\documentclass{llncs} 

\usepackage{version}
\usepackage{latexsym}

\begin{document} 

\newcommand{\labL}{{L_{\psi,E}}}
\newcommand{\bft}{\mbox{\em \bf true}}
\newcommand{\bff}{\mbox{\em \bf false}}
\newcommand{\zug}[1]{\langle #1  \rangle}
\newcommand{\cT}{{\cal T}}
\newcommand{\cO}{{\cal O}}
\newcommand{\bP}{{\bf P}}
\newcommand{\eqi}{\sim_i}

\newcommand\stt{{\rm \sigma}}

\newcommand\And{\wedge}
\newcommand\Or{\vee}
\newcommand\nxt{\bigcirc}

\newcommand\fin{{\it fin}}
\newcommand\init{{\it init}}
\newcommand\nll{\Lambda}

\newcommand\langc{{\cal L}_{n}^{C}}
\newcommand\langk{{\cal L}_{n}}

\newcommand{\vp}{\varphi}
\newcommand{\vpo}{\varphi}
\newcommand{\vptw}{\varphi}

\newcommand{\rarr}{\rightarrow} 

\newcommand{\lb}{\{} 
\newcommand{\rb}{\}} 
\newcommand{\view}{\{\cdot\}}
\newcommand{\viewpr}{\{\cdot\}^{pr}}

\newcommand{\ki}{K_i}

\newcommand{\pg}{{\bf Pg}} 
\newcommand{\pgi}{{\bf Pg}_i}
\newcommand{\pgj}{{\bf Pg}_j}

\newcommand{\act}{{\it ACT}}
\newcommand{\Prop}{{\it Prop}}
\newcommand{\goto}{{\rm goto}}
\newcommand{\assert}{{\rm assert}}

\newcommand{\ba}{{\bf a}} 
\newcommand{\ext}{{\bf ext}}
\newcommand{\bj}{{\bf j}}
\newcommand{\bp}{{\bf p}}

\newcommand{\cko}{{\cal K}_1}
\newcommand{\ckn}{{\cal K}_n}
\newcommand{\cki}{{\cal K}_i}
\newcommand{\join}{\bowtie}

\newcommand\obs{{\cal O}}
\newcommand{\clock}{clock} 

\newcommand\kbp{knowledge-based program}

\newcommand{\commentout}[1]{}
\commentout{
\long\def\ifundefined#1#2#3{\expandafter\ifx\csname#1\endcsname\relax
#2\else#3\fi}

\ifundefined{DEFNSTYPRESENT}{\def\thmcolon{\hspace{-.85em} {\bf
:} }}{\def\thmcolon{\relax}}

\newtheorem{THEOREM}{Theorem}[section]
\newenvironment{theorem}{\begin{THEOREM} \thmcolon }%
                        {\end{THEOREM}}
\newtheorem{LEMMA}[THEOREM]{Lemma}
\newenvironment{lemma}{\begin{LEMMA} \thmcolon }%
                      {\end{LEMMA}}
\newtheorem{COROLLARY}[THEOREM]{Corollary}
\newenvironment{corollary}{\begin{COROLLARY} \thmcolon }%
                          {\end{COROLLARY}}
\newtheorem{PROPOSITION}[THEOREM]{Proposition}
\newenvironment{proposition}{\begin{PROPOSITION} \thmcolon }%
                            {\end{PROPOSITION}}
\newtheorem{DEFINITION}[THEOREM]{Definition}
\newenvironment{definition}{\begin{DEFINITION} \thmcolon \rm}%
                            {\end{DEFINITION}}
\newtheorem{CLAIM}[THEOREM]{Claim}
\newenvironment{claim}{\begin{CLAIM} \thmcolon \rm}%
                            {\end{CLAIM}}
\newtheorem{EXAMPLE}[THEOREM]{Example}
\newenvironment{example}{\begin{EXAMPLE} \thmcolon \rm}%
                            {\end{EXAMPLE}}
\newtheorem{REMARK}[THEOREM]{Remark}
\newenvironment{remark}{\begin{REMARK} \thmcolon \rm}%
                            {\end{REMARK}}
}

\newcommand{\thm}{\begin{theorem}}
\newcommand{\lem}{\begin{lemma}}
\newcommand{\pro}{\begin{proposition}}
\newcommand{\dfn}{\begin{definition}}
\newcommand{\rem}{\begin{remark}}
\newcommand{\xam}{\begin{example}}
\newcommand{\cor}{\begin{corollary}}
\newcommand{\prf}{\noindent{\bf Proof:} }
\newcommand{\ethm}{\end{theorem}}
\newcommand{\elem}{\end{lemma}}
\newcommand{\epro}{\end{proposition}}
\newcommand{\edfn}{\bbox\end{definition}}
\newcommand{\erem}{\bbox\end{remark}}
\newcommand{\exam}{\bbox\end{example}}
\newcommand{\ecor}{\end{corollary}}
\newcommand{\eprf}{\bbox\vspace{0.1in}} 
\newcommand{\beqn}{\begin{equation}}
\newcommand{\eeqn}{\end{equation}}
\newcommand{\wbox}{\mbox{$\sqcap$\llap{$\sqcup$}}}
\newcommand{\bbox}{\vrule height7pt width4pt depth1pt}
\newcommand{\clm}{\begin{claim}}
\newcommand{\eclm}{\end{claim}}
\let\member=\in
\let\notmember=\notin
\newcommand{\sub}{_}
\def\su{^}
\newcommand{\rarrow}{\rightarrow}
\newcommand{\larrow}{\leftarrow}
\newcommand{\boldsymbol}[1]{\mbox{\boldmath $\bf #1$}}
\newcommand{\bolda}{{\bf a}}
\newcommand{\boldb}{{\bf b}}
\newcommand{\boldc}{{\bf c}}
\newcommand{\boldd}{{\bf d}}
\newcommand{\bolde}{{\bf e}}
\newcommand{\boldf}{{\bf f}}
\newcommand{\boldg}{{\bf g}}
\newcommand{\boldh}{{\bf h}}
\newcommand{\boldi}{{\bf i}}
\newcommand{\boldj}{{\bf j}}
\newcommand{\boldk}{{\bf k}}
\newcommand{\boldl}{{\bf l}}
\newcommand{\boldm}{{\bf m}}
\newcommand{\boldn}{{\bf n}}
\newcommand{\boldo}{{\bf o}}
\newcommand{\boldp}{{\bf p}}
\newcommand{\boldq}{{\bf q}}
\newcommand{\boldr}{{\bf r}}
\newcommand{\bolds}{{\bf s}}
\newcommand{\boldt}{{\bf t}}
\newcommand{\boldu}{{\bf u}}
\newcommand{\boldv}{{\bf v}}
\newcommand{\boldw}{{\bf w}}
\newcommand{\boldx}{{\bf x}}
\newcommand{\boldy}{{\bf y}}
\newcommand{\boldz}{{\bf z}}
\newcommand{\boldA}{{\bf A}}
\newcommand{\boldB}{{\bf B}}
\newcommand{\boldC}{{\bf C}}
\newcommand{\boldD}{{\bf D}}
\newcommand{\boldE}{{\bf E}}
\newcommand{\boldF}{{\bf F}}
\newcommand{\boldG}{{\bf G}}
\newcommand{\boldH}{{\bf H}}
\newcommand{\boldI}{{\bf I}}
\newcommand{\boldJ}{{\bf J}}
\newcommand{\boldK}{{\bf K}}
\newcommand{\boldL}{{\bf L}}
\newcommand{\boldM}{{\bf M}}
\newcommand{\boldN}{{\bf N}}
\newcommand{\boldO}{{\bf O}}
\newcommand{\boldP}{{\bf P}}
\newcommand{\boldQ}{{\bf Q}}
\newcommand{\boldR}{{\bf R}}
\newcommand{\boldS}{{\bf S}}
\newcommand{\boldT}{{\bf T}}
\newcommand{\boldU}{{\bf U}}
\newcommand{\boldV}{{\bf V}}
\newcommand{\boldW}{{\bf W}}
\newcommand{\boldX}{{\bf X}}
\newcommand{\boldY}{{\bf Y}}
\newcommand{\boldZ}{{\bf Z}}
\newcommand{\sat}{\models}
\newcommand{\dtur}{\models}
\newcommand{\infers}{\vdash}
\newcommand{\stur}{\vdash}
\newcommand{\rimp}{\Rightarrow}
\newcommand{\limp}{\Leftarrow}
\newcommand{\dimp}{\Leftrightarrow}
\newcommand{\bor}{\bigvee}
\newcommand{\band}{\bigwedge}
\newcommand{\union}{\cup}
\newcommand{\inter}{\cap}
\newcommand{\xx}{{\bf x}}
\newcommand{\yy}{{\bf y}}
\newcommand{\uu}{{\bf u}}
\newcommand{\vv}{{\bf v}}
\newcommand{\FF}{{\bf F}}
\newcommand{\natnum}{{\sl N}}
\newcommand{\IR}{\mbox{$I\!\!R$}}
\newcommand{\IP}{\mbox{$I\!\!P$}}
\newcommand{\IN}{\mbox{$I\!\!N$}}
\newcommand{\IC}{\mbox{$C\!\!\!\!\raisebox{.75pt}{\mbox{\sqi I}}$}}
\newcommand{\marrow}{\hbox{$\rightarrow$ \hskip -10pt
                      $\rightarrow$ \hskip 3pt}}
\renewcommand{\phi}{\varphi}
\newcommand{\Circ}{\mbox{{\small $\bigcirc$}}}
\newcommand{\lt}{<}
\newcommand{\gt}{>}
\newcommand{\all}{\forall}
\newcommand{\infinity}{\infty}
\newcommand{\bc}[2]{\left( \begin{array}{c} #1 \\ #2 \end{array} \right)}
\newcommand{\cross}{\times}
\newcommand{\bigfootnote}[1]{{\footnote{\normalsize #1}}}
\newcommand{\medfootnote}[1]{{\footnote{\small #1}}}
\newcommand{\bd}{\bf}

\newcommand{\imp}{\Rightarrow}

\newcommand{\A}{{\cal A}}
\newcommand{\B}{{\cal B}}
\newcommand{\C}{{\cal C}}
\newcommand{\D}{{\cal D}}
\newcommand{\E}{{\cal E}}
\newcommand{\F}{{\cal F}}
\newcommand{\G}{{\cal G}}
\newcommand{\I}{{\cal I}}
\newcommand{\J}{{\cal J}}
\newcommand{\K}{{\cal K}}
\newcommand{\M}{{\cal M}}
\newcommand{\N}{{\cal N}}
\newcommand{\Ocal}{{\cal O}}
\newcommand{\Hcal}{{\cal H}}
\renewcommand{\P}{{\cal P}}
\newcommand{\Q}{{\cal Q}}
\newcommand{\R}{{\cal R}}
\newcommand{\T}{{\cal T}}
\newcommand{\U}{{\cal U}}
\newcommand{\V}{{\cal V}}
\newcommand{\W}{{\cal W}}
\newcommand{\X}{{\cal X}}
\newcommand{\Y}{{\cal Y}}
\newcommand{\Z}{{\cal Z}}

\newcommand{\Kone}{{\cal K}_1}
\newcommand{\abs}[1]{\left| #1\right|}
\newcommand{\set}[1]{\left\{ #1 \right\}}
\newcommand{\Ki}{{\cal K}_i}
\newcommand{\Kn}{{\cal K}_n}
\newcommand{\st}{\, \vert \,} %
\newcommand{\stc}{\, : \,} %
\newcommand{\la}{\langle}
\newcommand{\ra}{\rangle}
\newcommand{\<}{\langle}
\renewcommand{\>}{\rangle}
\newcommand{\lang}{\mbox{${\cal L}_n$}}
\newcommand{\langd}{\mbox{${\cal L}_n^D$}}

\newtheorem{nlem}{Lemma}
\newtheorem{Ob}{Observation}
\newtheorem{pps}{Proposition}
\newtheorem{defn}{Definition}
\newtheorem{crl}{Corollary}
\newtheorem{cl}{Claim}
\newcommand{\pf}{\par\noindent{\bf Proof}~~}
\newcommand{\eg}{e.g.,~}
\newcommand{\ie}{i.e.,~}
\newcommand{\vs}{vs.~}
\newcommand{\cf}{cf.~}
\newcommand{\etal}{et al.\ }
\newcommand{\resp}{resp.\ }
\newcommand{\respc}{resp.,\ }
\newcommand{\wrt}{with respect to~}
\newcommand{\re}{r.e.}
\newcommand{\nind}{\noindent}
\newcommand{\distributed}{distributed\ }
\newcommand{\bn}{\bigskip\markright{NOTES}
\section*{Notes}}
\newcommand{\Exer}{
\bigskip\markright{EXERCISES}
\section*{Exercises}}
\newcommand{\DG}{D_G}
\newcommand{\Sm}{{\rm S5}_m}
\newcommand{\Smc}{{\rm S5C}_m}
\newcommand{\Smi}{{\rm S5I}_m}
\newcommand{\Smic}{{\rm S5CI}_m}
\newcommand{\Martin}{Mart\'\i n\ }
\newcommand{\ol}{\setlength{\itemsep}{0pt}\begin{enumerate}}
\newcommand{\eol}{\end{enumerate}\setlength{\itemsep}{-\parsep}}
\newcommand{\ul}{\setlength{\itemsep}{0pt}\begin{itemize}}
\newcommand{\dl}{\setlength{\itemsep}{0pt}\begin{description}}
\newcommand{\edl}{\end{description}\setlength{\itemsep}{-\parsep}}
\newcommand{\eul}{\end{itemize}\setlength{\itemsep}{-\parsep}}
\newtheorem{fthm}{Theorem}
\newtheorem{flem}[fthm]{Lemma}
\newtheorem{fcor}[fthm]{Corollary}
\newcommand{\slidehead}[1]{
\eject
\Huge
\begin{center}
{\bf #1 }
\end{center}
\vspace{.5in}
\LARGE}

\newcommand{\subG}{_G}
\newcommand{\If}{{\bf if}}

\newcommand{\attime}{{\tt \ at\_time\ }}
\newcommand{\hatell}{\skew6\hat\ell\,}
\newcommand{\Then}{{\bf then}}
\newcommand{\Until}{{\bf until}}
\newcommand{\Else}{{\bf else}}
\newcommand{\Repeat}{{\bf repeat}}
\newcommand{\cA}{{\cal A}}
\newcommand{\cL}{{\cal L}}
\newcommand{\cE}{{\cal E}}
\newcommand{\cF}{{\cal F}}
\newcommand{\cI}{{\cal I}}
\newcommand{\cN}{{\cal N}}
\newcommand{\cR}{{\cal R}}
\newcommand{\cS}{{\cal S}}
\newcommand{\BN}{B^{\scriptscriptstyle \cN}}
\newcommand{\BS}{B^{\scriptscriptstyle \cS}}
\newcommand{\cW}{{\cal W}}
\newcommand{\EG}{E_G}
\newcommand{\CG}{C_G}
\newcommand{\CN}{C_\cN}
\newcommand{\ES}{E_\cS}
\newcommand{\EN}{E_\cN}
\newcommand{\CS}{C_\cS}

\newcommand{\attack}{\mbox{{\it attack}}}
\newcommand{\attacking}{\mbox{{\it attacking}}}
\newcommand{\delivered}{\mbox{{\it delivered}}}
\newcommand{\exist}{\mbox{{\it exist}}}
\newcommand{\decide}{\mbox{{\it decide}}}
\newcommand{\clean}{{\it clean}}
\newcommand{\diff}{{\it diff}}
\newcommand{\Failed}{{\it failed}}
\newcommand\eqdef{=_{\rm def}}
\newcommand{\true}{\mbox{{\it true}}}
\newcommand{\false}{\mbox{{\it false}}}

\newcommand{\DN}{D_{\cN}}
\newcommand{\DS}{D_{\cS}}
\newcommand{\tyme}{{\it time}}
\newcommand{\fp}{f}

\newcommand{\Kax}{{\rm K}_n}
\newcommand{\Kaxc}{{\rm K}_n^C}
\newcommand{\Kaxd}{{\rm K}_n^D}
\newcommand{\Tax}{{\rm T}_n}
\newcommand{\Taxc}{{\rm T}_n^C}
\newcommand{\Taxd}{{\rm T}_n^D}
\newcommand{\fourax}{{\rm S4}_n}
\newcommand{\fouraxc}{{\rm S4}_n^C}
\newcommand{\fouraxd}{{\rm S4}_n^D}
\newcommand{\fiveax}{{\rm S5}_n}
\newcommand{\fiveaxc}{{\rm S5}_n^C}
\newcommand{\fiveaxd}{{\rm S5}_n^D}
\newcommand{\Dax}{{\rm KD45}_n}
\newcommand{\Daxc}{{\rm KD45}_n^C}
\newcommand{\Daxd}{{\rm KD45}_n^D}
\newcommand{\LP}{{\cal L}_n}
\newcommand{\LCP}{{\cal L}_n^C}
\newcommand{\LDP}{{\cal L}_n^D}
\newcommand{\LCDP}{{\cal L}_n^{CD}}
\newcommand{\MP}{{\cal M}_n}
\newcommand{\MPr}{{\cal M}_n^r}
\newcommand{\MPrt}{\M_n^{\mbox{\scriptsize{{\it rt}}}}}
\newcommand{\MPrst}{\M_n^{\mbox{\scriptsize{{\it rst}}}}}
\newcommand{\MPelt}{\M_n^{\mbox{\scriptsize{{\it elt}}}}}
\renewcommand{\lang}{\mbox{${\cal L}_{n} (\Phi)$}}
\renewcommand{\langd}{\mbox{${\cal L}_{n}^D (\Phi)$}}
\newcommand{\fiveaxdu}{{\rm S5}_n^{DU}}
\newcommand{\LPD}{{\cal L}_n^D}
\newcommand{\fiveaxu}{{\rm S5}_n^U}
\newcommand{\fiveaxcu}{{\rm S5}_n^{CU}}
\newcommand{\LPU}{{\cal L}^{U}_n}
\newcommand{\LPCU}{{\cal L}_n^{CU}}
\newcommand{\LDPU}{{\cal L}_n^{DU}}
\newcommand{\LCPU}{{\cal L}_n^{CU}}
\newcommand{\LPDU}{{\cal L}_n^{DU}}
\newcommand{\LPCDU}{{\cal L}_n^{\it CDU}}
\newcommand{\Cn}{\C_n}
\newcommand{\CSnp}{\I_n^{oa}(\Phi')}
\newcommand{\CSc}{\C_n^{oa}(\Phi)}
\newcommand{\Ccs}{\C_n^{oa}}
\newcommand{\CSAX}{OA$_{n,\Phi}$}
\newcommand{\CSAXN}{OA$_{n,{\Phi}}'$}
\newcommand{\untill}{U}
\newcommand{\until}{\, U \,}
\newcommand{\amp}{{\rm a.m.p.}}

\newcommand{\msgc}[1]{ @ #1 }
\newcommand{\Camp}{{\C_n^{\it amp}}}
\newcommand{\bi}{\begin{itemize}}
\newcommand{\ei}{\end{itemize}}
\newcommand{\be}{\begin{enumerate}}
\newcommand{\ee}{\end{enumerate}}
\newcommand{\rarrowr}{\stackrel{r}{\rightarrow}}
\newcommand{\ack}{\mbox{\it ack}}
\newcommand{\Gz}{\G_0}
\newcommand{\denselist}{\itemsep 0pt\partopsep 0pt}
\def\seealso#1#2{({\em see also\/} #1), #2}

\newcommand{\said}{{\rm said}}

\newcommand{\pow}[1]{\P(#1)}

{\small 

\title{Synthesis from Knowledge-Based Specifications\thanks{
An extended abstract of this paper appeared in CONCUR'98. 
Work begun 
while both authors were visitors at the DIMACS Special
Year on Logic and Algorithms. Work of the first author supported by an
Australian Research Council Large Grant. 
Work of the second author supported in part  
by NSF grants CCR-9628400 and 
CCR-9700061, and by a grant from the Intel Corporation.
Thanks to Kai Engelhardt, 
Yoram Moses and Nikolay Shilov
for their 
comments on earlier versions of this paper.
}
} 

\date{}
\titlerunning{Synthesis from Knowledge-Based Specifications}  %
\author{Ron van der Meyden\inst{1} \and Moshe Y. Vardi\inst{2}
}
\authorrunning{van der Meyden, Vardi}   %
\tocauthor{Ron van der Meyden (University of Technology, Sydney), 
Moshe Y. Vardi (Rice)}
\institute{School of Computer Science and Engineering \\ 
University of New South Wales\\ 
Sydney, NSW 2021, Australia\\ 
{\tt meyden@cse.unsw.edu.au}\\
{\tt http://www.cse.unsw.edu.au/$^\sim$meyden}\\
\and
Department of Computer Science \\
    Mail Stop 132,
    Rice University \\ 
    6100 S. Main Street \\ 
    Houston, TX 77005-1892, U.S.A\\ 
{\tt vardi@cs.rice.edu}\\
{\tt http://www.cs.rice.edu/$^\sim$vardi}}

\maketitle

\begin{abstract}
In program synthesis, we transform a specification into a program that
is guaranteed to satisfy the specification.  In synthesis of reactive
systems, the environment in which the program operates may behave
nondeterministically, e.g., by generating different sequences of
inputs in different runs of the system. To satisfy the specification,
the program needs to act so that the specification holds in every
computation generated by its interaction with the environment.  Often,
the program cannot observe all attributes of its environment.  In this
case, we should transform a specification into a program whose
behavior depends only on the observable history of the
computation. This is called {\em synthesis with incomplete
information\/}.  In such a setting, it is desirable to have a {\em
knowledge-based specification\/}, which can refer to the uncertainty
the program has about the environment's behavior.  In this work we
solve the problem of synthesis with incomplete information with
respect to specifications in the logic of knowledge and time. We show
that the problem has the same worst-case complexity as synthesis with
complete information.

\end{abstract}

\section{Introduction}

One of the most significant developments in the area of design
verification 
is the development of
of algorithmic methods for verifying temporal specifications
of {\em finite-state\/} designs 
\cite{CES86,LP85,QS81,VW86b}.
The significance of this follows from the fact that a
considerable number of the communication and synchronization
protocols studied in the literature are in essence finite-state
programs or can be abstracted as finite-state programs.
A frequent criticism against this approach, however, is that 
verification is done {\em after\/} substantial resources have 
already been invested in the development of the design.
Since designs invariably contain errors, verification simply
becomes part of the debugging process.
The critics argue that the desired goal is to use the specification
in the design development process in order to guarantee the
design of correct programs.  This is called {\em program synthesis}.

The classical approach to program synthesis is to extract a program
from a proof that the specification is satisfiable.
For reactive programs, the specification is typically a temporal
formula describing the allowable behaviors of the program
\cite{MP92}.
Emerson and Clarke \cite{EC82} and Manna and Wolper \cite{MW84}
showed how to extract programs from 
(finite representations of) 
models of the formula.
In the late 1980s, several researchers realized that the classical
approach is well suited to {\em closed\/} systems, but not to
{\em open\/} systems \cite{Dil89,PR89a,ALW89}.
In open systems the program interacts with the environment.
A correct program should be able to handle arbitrary actions of the
environment.
If one applies the techniques of \cite{EC82,MW84} to open systems,
one obtains programs that can handle only some actions of the 
environment.

Pnueli and Rosner \cite{PR89a}, Abadi, Lamport and Wolper 
\cite{ALW89}, and Dill \cite{Dil89}
argued that the right way to approach synthesis of open systems
is to consider the situation as a (possibly infinite) game between
the environment and the program.
A correct program 
can then be viewed 
as a winning strategy in this game.
It turns out that satisfiability of the specification is not
sufficient to guarantee the existence of such a strategy.
Abadi et al. called specifications for which winning strategies
exist {\em realizable}.
A winning strategy can be viewed as an infinite tree.
In those papers it is shown how the specification can be transformed
into a tree automaton such that a program is
realizable precisely when this tree automaton is nonempty, i.e.,
it accepts some infinite tree.
This yields a decision procedure for realizability.
(This is closely related to the approach taken by B\"uchi and 
Landweber \cite{BL69} and Rabin \cite{Rab70}
to solve Church's {\em solvability problem} \cite{Chu63}.)

The works discussed so far deal with situations in which the
program has complete information about
the actions taken by the environment.
This is called synthesis with {\em complete information}. 
Often, the program does not have complete information about its
environment. Thus, the actions of the program can depend only
on the ``visible'' part of the computation.
Synthesizing such programs is called synthesis with
{\em incomplete information}.
The difficulty of synthesis with incomplete information follows
from the fact that while in the complete-information case the
strategy tree and the computation tree coincide, this is no longer
the case when we have incomplete information.
Algorithms for synthesis were extended to handle 
incomplete information in \cite{PR89b,WD90,AM94,KS95,Var95c,KV97c}.

It is important to note that temporal logic specifications
cannot refer to the uncertainty of the program about the
environment, since the logic has no construct for referring to
such uncertainty.
It has been observed, however, that designers of open systems often
reason explicitly in terms of uncertainty \cite{Hal1}.
A typical example is a rule of the form ``send an acknowledgement
as soon as you {\em know\/} that the message has been received''.
For this reason, it has been proposed in \cite{HM90}
to use epistemic logic as a specification language for open 
systems with incomplete information.
When dealing with ongoing behavior in systems with incomplete
information, a combination of temporal and epistemic logic
can refer to both behavior and uncertainty \cite{Leh,LR}.
In such a logic the above rule can be formalized by the formula
$\Box (K {\sf received} \rightarrow {\sf ack})$, where $\Box$ is the
temporal connective ``always'', $K$ is the epistemic modality
indicating knowledge, and ${\sf received}$ and ${\sf ack}$ are
atomic propositions.

Reasoning about open systems at the {\em knowledge level\/} 
allows us to abstract away from many concrete
details of the systems we are considering.
It is often more intuitive to think in terms of the high-level
concepts when we design a protocol, and then translate these
intuitions into a concrete program, based on the particular
properties of the setting we are considering.
This style of program development
will generally allow us to modify the program more easily when
considering a setting with different properties, such as different
communication topologies, different guarantees about the reliability
of various components of the system, and the like.
See \cite{APP,BrafmanLMS1997,ChM,DM,Had,HMW,HZ,MT,NeigerBazzi,%
NeigTuttle,Rub89}
for examples of knowledge-level analysis of open systems with
incomplete information.
To be able to translate, however, these high-level intuitions into
a concrete program one has to be able to check that the given
specification is realizable in the sense described above.

Our goal in this paper is to extend the program synthesis framework
to temporal-epistemic specification.
The difficulty that we face is that all previous program-synthesis
algorithms attempt to construct strategy trees that realize the given
specification. Such trees, however, refer to temporal behavior
only and they do not contain enough information to interpret the
epistemic constructs.
(We note that this difficulty is different than the difficulty
faced when one attempts to extend synthesis with
incomplete information to {\em branching-time\/} specification
\cite{KV97c}, and the solution described there cannot be applied
to knowledge-based specifications.)
Our key technical tool is the definition of {\em finitely labelled}
trees that contain information about both temporal behavior 
and epistemic uncertainty. 
Our main result is that 
we can extend the program synthesis framework to handle
knowledge-based specification with no increase in worst-case
computational complexity.

In an earlier, extended abstract of the present work \cite{MV98} we stated
this result for specifications in the logic of knowledge and
{\em linear} time, and required the protocols synthesized to be deterministic. 
The present paper differs from the earlier work in giving
full proofs of all results, as well as in the fact that we generalize the 
specification language to encompass branching as well as linear time logical 
operators. We also liberalize the class of solutions to encompass nondeterministic 
protocols. (Our previous result on deterministic protocols is easily recovered, 
by noting that the branching time specification language can express determinism of the solutions.) 
These generalizations allow us to give an application of the results to the synthesis of 
implementations of knowledge-based programs \cite{FHMV}, 
a type of programs in an agent's actions may depend in its knowledge. 

The structure of the paper is as follows. Section~\ref{sec:defs} defines the syntax and 
semantics of the temporal-epistemic specification language and defines the synthesis
problem for this language. In Section~\ref{sec:charn} we give a characterization of 
realizability that forms the basis for our synthesis result. 
Section~\ref{sec:alg} describes an automaton-theoretic algorithm for 
deciding whether a specification is realizable, and for extracting a
solution in case it is. Section~\ref{sec:discn} discusses two aspects of this 
result: a subtlety concerning the knowledge encoded in the states of the solutions, 
and an application of our result to knowledge-based program implementation. 
The paper concludes in Section~\ref{sec:concl} with a discussion of 
extensions and open problems.

\section{Definitions} \label{sec:defs}

In this section we define the formal framework within which we will
study the problem of synthesis from knowledge-based specifications,
provide semantics for the logic of knowledge and time in this
framework, and define the realizability problem. 

Systems will be decomposed in our framework into two components: the
program, or \emph{protocol} being run, 
and the
remainder of the system, which we call the \emph{environment} within
which this protocol operates.  We begin by presenting a model, from
\cite{meyden:tark96}, for the environment. This model is an adaption 
of the notion of {\em context} of Fagin et al.~\cite{FHMV}.
Our main result in this paper is restricted to the case of a single 
agent, but as we will state a result in Section~\ref{sec:kii} that 
applies in a more general setting, we define the model assuming a 
finite number of agents.  

Intuitively, we model the environment as a finite-state transition 
system, with the transitions labelled by the agents' actions.
For each agent $i=1\ldots n$ let $\act_i$ be a set of
\emph{actions} associated with agent $i$. We will also consider the
environment as able to perform actions, so assume additionally a set 
$\act_e$ of actions for the environment.  A \emph{joint action} will
consist of an action for each agent and an action for the environment,
i.e., the set of joint actions is the cartesian product 
$\act=
\act_e \times \act_1\times \ldots \times \act_n$.

Suppose we are given such a set of actions, together with a set of
$\Prop$ of atomic propositions. Define a \emph{finite interpreted
environment for $n$ agents} to be a tuple $E$ of the form $\la S_e,
I_e, P_e, \tau , O_1,\ldots, O_n,\pi_e \ra$ where the components are as
follows:
\begin{enumerate}

\item $S_e$ is a finite set of {\em
states of the environment}. Intuitively, states of the environment may
encode such information as messages in transit, failure of components,
etc.  and possibly the values of certain local variables maintained by
the agents.

\item  $I_e$ is a subset of $S_e$, representing the possible 
{\em initial states\/} of the environment. 

\item $P_e:S_e \rightarrow {\cal P}(\act_e)$ is a function,
called the {\em protocol of the environment,} mapping 
states to subsets of the set $\act_e$ of actions performable by the
environment.  Intuitively, $P_e(s)$ represents the set of actions that
may be performed by the environment when the system is in state $s$.
We assume that this set is nonempty for all $s\in S_e$. 
 
\item $\tau$ is a function mapping joint actions ${\bf a}\in \act$ to 
state transition functions $\tau({\bf a}):S_e \rightarrow S_e$. 
Intuitively, when the joint action ${\bf a}$ is performed
in the state $s$, the resulting state of  the environment is 
$\tau({\bf a})(s)$. 

\item For each $i=1\ldots n$, the component $O_i$ is a function, called the
{\em observation function of agent $i$,} mapping the set of states
$S_e$ to some set ${\cal O}$.  If $s$ is a global state then $O_i(s)$
will be called the {\em observation} of agent $i$ in the 
state $s$.

\item $\pi_e:S_e\rightarrow\{0,1\}^\Prop$ is an \emph{interpretation}, 
mapping each state to an assignment of truth values
to the atomic propositions in $\Prop$. 

\end{enumerate}

A {\em run} 
$r$
of an environment $E$ is an {\em infinite} sequence 
$s_0, s_1,\ldots $ 
of states such that $s_0\in I_e$ and for all $m\geq 0$
there exists a joint action ${\bf a}= \la a_e, a_1,\ldots, a_n\ra$
such that $s_{m+1} = \tau({\bf a})(s_m)$ and $a_e \in P_e(s_m)$. For
$m\geq 0$ we write 
$r(m)$
for $s_m$. For $k\leq m$ we also write
$r[k..m]$ for the sequence $s_k\ldots s_m$ and $r[m..]$ for $s_m
s_{m+1}\ldots$.

A \emph{point} is a tuple $(r,m)$, where $r$ is a run and $m$ a
natural number. Intuitively, a point identifies a particular instant
of time along the history described by the run. A run $r'$ will be
said to be a run \emph{through} a point $(r,m)$ if $r[0..m] =
r'[0..m]$. Intuitively, this is the case when the two runs $r$ and $r'$
describe the same sequence of events up to time $m$.

Runs of an environment provide sufficient structure for the
interpretation of formulae of linear temporal logic. To interpret
formulae involving knowledge, we need additional structure. Knowledge
arises not from a single run, but from the position a run occupies
within 
the collection of all possible runs of the system under study. 
Following \cite{FHMV}, define a \emph{system} to be a set $\cR$ of
runs and an \emph{interpreted system} to be a tuple $\cI = (\cR, \pi)$
consisting of a system $\cR$ together with an interpretation function
$\pi$ mapping the points of runs in $\cR$ to assignments of truth
value to the propositions in $\Prop$.  
As we will show below,
interpreted systems also provide enough structure to interpret
branching time logics 
\cite{CES86},
 by means of a slight modification of
the usual semantics for such logics.

All the interpreted systems we deal with in this paper will have all
runs drawn from the same environment, and the interpretation $\pi$
derived from the interpretation of the environment by means of the
equation $\pi(r,m)(p) = \pi_e(r(m))(p)$, where $(r,m)$ is a point
and $p$ an atomic proposition. That is, the value of a proposition at
a point of a run is determined from the state of the environment at
that point, as described by the environment generating the run.

The definition of run presented above is a slight modification of the
definitions of Fagin et al. \cite{FHMV}. Roughly corresponding to 
our notion of state of the environment is their  notion of a \emph{global
state}, which has additional structure. Specifically, a global state
identifies a \emph{local state} for each agent, which plays a crucial
role in the semantics of knowledge. We have avoided the use of such
extra structure in our states because we focus on just one particular 
definition of local states that may be represented in the general
framework of \cite{FHMV}.

In particular, we will work with respect to a 
\emph{synchronous perfect-recall}
semantics of knowledge. Given a run $r= s_0,s_1 \ldots$ of an
environment with observation functions $O_i$, we define the {\em local
state of agent $i$ at time $m\geq 0$} to be the sequence $r_i(m) =
O_i(s_0)\ldots O_i(s_m)$.  That is, the local state of an agent at a
point in a run consists of a complete record of the observations the
agent has made up to that point.

These local states may be used to define for each agent $i$ a relation
$\eqi$ of \emph{indistinguishability} on points, by $(r,m) \eqi
(r',m')$ if $r_i(m) = r'_i(m')$. Intuititively, when $(r,m) \eqi
(r',m')$, agent $i$ has failed to receive enough information to time
$m$ in run $r$ and time $m'$ in run $r'$ to determine whether it is on
one situation or the other.  Clearly, each $\eqi$ is an equivalence
relation.  The use of the term ``synchronous'' above is due to the
fact that an agent is able to determine the time simply by counting
the number of observations in its local state. This is reflected in
the fact that if $(r,m) \eqi (r',m')$, we must have $m=m'$. 
(There
also exists an \emph{asynchronous} version of perfect recall 
\cite{FHMV},
which will not concern us in the present paper.)

To specify systems, we will use a propositional multimodal language
for knowledge and 
time based on a set $\Prop$ of atomic
propositions, with formulae generated by the modalities $\nxt$ (next
time), $\until$ (until), and a knowledge operator $K_i$ for each agent
$i=1\ldots n$. 
Time may be either branching or linear, so we also
consider the branching time quantifier $\exists$. 
More precisely, the
set of formulae of the language is defined as follows: each atomic
proposition $p\in \Prop$ is a formula, and if $\vp$ and $\psi$ are
formulae, then so are $\neg \vp$, $\vp\And\psi$, $\nxt
\vp$, $\vp \until\psi$, 
$\exists \phi$ and 
$K_i\vp$ for each $i=1\ldots n$. 
As usual, we use the abbrevations $\Diamond \vp$ for 
${\bf true} \until \vp$, and $\Box\vp$ for $\neg \Diamond \neg \vp$.

The semantics of this language is defined as follows. Suppose we are
given an interpreted system $\cI = (\cR,\pi)$, where $\cR$ is a set of
runs of environment $E$ and $\pi$ is determined from the environment
as described above.  We define satisfaction of a formula $\vp$ at a
point $(r,m)$ of a run in $\cR$, denoted $\cI,(r,m)\models \vp$,
inductively on the structure of $\vp$. The cases for the temporal
fragment of the language are standard:
\be
\item $\cI,(r,m)\models p$, where $p$ is an atomic proposition, if 
$\pi(r,m)(p)= 1$, 
\item $\cI,(r,m)\models \vp_1 \And \vp_2$, if 
$\cI,(r,m)\models \vp_1$ and $\cI,(r,m)\models \vp_2$, 
\item $\cI,(r,m)\models \neg \vp$, if not 
$\cI,(r,m)\models \vp$, 
\item $\cI,(r,m)\models \nxt \vp$, if 
$\cI,(r,m+1)\models \vp$, 
\item $\cI,(r,m)\models \vp_1\until \vp_2$, if there 
exists $k\geq m$ such that 
$\cI,(r,k)\models \vp_2$ and $\cI,(r,l)\models \vp_1$
for all $l$ with $m \leq l <k$. 

\item $\cI,(r,m)\models \exists \vp$ if there exists a 
run $r'$ in $\cR$ through $(r,m)$ such that $\cI,(r',m)\models \vp$.

\ee 

The semantics of the knowledge operators is defined by 
\be
\item[7.] $\cI,(r,m)\models K_i\vp$, if 
$\cI,(r',m')\models \vp$ for all points $(r',m')$ of $\cI$ 
satisfying $(r',m') \eqi (r,m)$
\ee
That is, an agent knows a formula to be true if this formula
holds at all points that it is unable to distinguish from the 
actual point. This definition follows the general framework for the 
semantics of knowledge proposed by Halpern and Moses \cite{HM90}. 
We use the particular equivalence relations $\eqi$ 
obtained from the assumption of synchronous perfect recall, but the 
same semantics for knowledge applies for other ways of defining
local states, and hence the relations $\eqi$. We refer the reader to 
\cite{HM90,FHMV} for further background on this topic.

The systems $\cI$ we will be interested in will not have completely
arbitrary sets of runs, but rather will have sets of runs that arise
from the agents running some program, or protocol, within a given
environment.
Intuitively, an agent's 
choice of 
actions in such a program should depend on the
information it has been able to obtain about the environment, but no
more.  We have used observations to model the agent's source of
information about the environment. The maximum information that an
agent has  about the environment at a point $(r,m)$ is given
by the local state $r_i(m)$. Thus, it is natural to 
model an agent's program as assigning to each local state
of the agent 
a nonempty set of actions
for that agent. 
We define a \emph{protocol} 
for agent $i$ to be a function 
$P_i:\cO^+ \rarr \pow{ACT_i}\setminus \{\emptyset\}$. 
A {\em joint protocol\/} $\bP$ is a tuple $\la P_1,\ldots,P_n\ra$, where 
each $P_i$ is a protocol for agent $i$. 
We say that $\bP$ is {\em deterministic} if $\P_i(v)$ is a singleton for all agents $i$ and local states $v$.

The systems we consider will consist of all the runs in which at each
point of time each agent behaves as required by its protocol. As
usual, we also require that the environment follows its own protocol.
Formally, the {\em system generated by a joint protocol $\bP$ in
environment $E$} is the set $\cR(\bP,E)$ of all runs $r$ of $E$ such
that for all $m\geq 0$ we have $r(m+1)= \tau(\ba)(r(m))$, where
$\ba$ is a joint action in $P_e(r(m))\times P_1(r_1(m))\times \ldots \times P_n(r_n(m))$.  
The {\em interpreted system generated by a joint
protocol $\bP$ in environment $E$} is the interpreted system
$\cI(\bP,E) = (\cR(\bP,E),
\pi)$, where $\pi$ is the interpretation derived from the environment
$E$ as described above.

Finally, we may define the relation between specifications and
implementations that is our main topic of study.  We say that a joint 
protocol $\bP$ {\em realizes} a specification $\vp$ in an environment
$E$ if for all runs $r$ of $\cI(\bP,E)$ we have 
$\cI(\bP,E),(r,0)
\models \vp $. A specification $\vp$ is {\em realizable} in
environment $E$ if there exists a joint protocol $\bP$ that realizes
$\vp$ in $E$. The following example illustrates the framework 
and provides examples of realizable and unrealizable formulae.

\begin{example} \label{ex:tog} 
Consider a timed toggle switch with two positions (on, off), with a
light intended to indicate the position. If the light is on, then the
switch must be in the on position. However, the light is faulty, so it
might be off when the switch is on.  Suppose that there is a single
agent that has two actions: 
``toggle" and ``do nothing". 
If the agent
toggles, the switch changes position. If the agent does nothing, the
toggle either stays in the same position or, if it is on, may timeout
and switch to off automatically.  The timer is unreliable, so the
timeout may happen any time the switch is on, or never, even if the
switch remains on forever.  The agent observes only the light, not the
toggle position. 

This system may be represented as an environment with states
consisting of pairs $\la t,l\ra$, where $t$ is a boolean variable
indicating the toggle position and $l$ is a boolean variable
representing the light, subject to the constraint
that $l=0$ if $t=0$. The agent's observation function is given by
$O_1(\la t,l\ra) = l$.  To represent the effect of the agent's actions
on the state, write $T$ for the toggle action and $\Lambda$ for the
agent's null action.  The environment's actions may be taken to be
pairs $(u,v)$ where $u$ and $v$ are boolean variables indicating,
respectively, that the environment times out the toggle, and that it
switches the light on (provided the switch is on). Thus, the transition
function is given by $\tau(\la (u,v),a_1 \ra)(\la t,l\ra) = \la
t',l'\ra$ where (i) $t' =
\overline{t}$ if either $a_1 = T$ or $t=u=1$, else $t'=t$, and (ii)
$l'=1$ iff $t'=1$ and $v=1$.

If ``$\mbox{\tt toggle-on}$'' is the proposition true in states $\la t,l\ra$ where
$t=1$, then the formula $\Box ( K_1 \mbox{\tt toggle-on} \Or K_1 \neg
\mbox{\tt toggle-on})$ expresses that the agent knows at all times whether or
not the toggle is on. This formula is realizable when the initial
states of the environment are those in which the toggle is on (and the
light is either on or off).  The protocol by which the agent realizes
this formula is that in which it performs $T$ at all steps.  Since it has
perfect recall it can determine whether the toggle is on or off by
checking if it has made (respectively) an odd or an even number of
observations. 

However, the same formula is not realizable if all states are
initial. In this case, if the light is off at time 0, the agent cannot
know whether the switch is on.  As it has had at time 0 no opportunity
to influence the state of the environment through its actions, this is
the case whatever the agent's protocol. $\Box$ 
\end{example}

\section{A Characterization of Realizability} \label{sec:charn}

In this section we characterize realizability in environments for a
single agent in terms of the existence of a certain type of labelled
tree.
Intuitively, the nodes of this tree correspond to the local states
of the agent, and the label at a node  is intended to express (i) the
relevant knowledge of the agent and (ii) the action the agent performs
when in the corresponding local state.

Consider $\cO^*$, the set of all 
finite sequences of observations
of agent 1, including the empty sequence. 
This set may be viewed as an infinite tree, where the root is the 
null sequence and the successors of a vertex
$v\in \cO^*$ are the vertices $v\cdot o$, where $o\in \cO$ is an 
observation.
A {\em labelling}  of $\cO^*$ is a function $\cT:\cO^* \rarr L$ for 
some set $L$. We call $\cT$ a {\em labelled tree}.
We will work with trees in which the labels are constructed from the
states of the environment, a formula $\psi$ and the actions of the
agent.  Define an {\em atom\/} for a formula $\psi$ to be a mapping
$X$ from the set of all subformulae of $\psi$ to $\{0,1\}$.  A {\em
knowledge set\/} for $\psi$ in $E$ is a set of pairs of the form
$(X,s)$, where $X$ is an atom of $\psi$ and $s$ is a state of $E$.
Take 
$\labL$
to be the set of all pairs of the form 
$(K,B)$ 
where $K$ is
a knowledge set for $\psi$ in $E$ and 
$B\subseteq \act_1$ is a nonempty set of actions
of agent 1.  We
will consider trees that are labellings of $\cO^*$ by 
$\labL$.
We will call such a tree a {\em labelled tree for $\psi$ and $E$.}

Given such a labelled tree $\cT$, we may define the functions $K$,
mapping $\cO^*$ to knowledge sets, and $P$, mapping $\cO^*$ to 
nonempty sets of 
actions
of agent 1, such that for all $v\in \cO^*$ we have $\cT(v) =
(K(v),P(v))$.  Note that $P$ is a protocol for agent 1.
This protocol generates an interpreted system $\cI(P,E)$ in the given
environment $E$.  
Intuitively, we are interested in trees in which the $K(v)$ 
describe the states of knowledge of the agent in this system.
We now set about stating some
constraints on the labels in the tree $\cT$ that are intended to
ensure 
this is the case. 

Suppose we are given a sequence of states $r = s_0s_1\ldots$ and a vertex
$v$ of $\cT$ with $v= w\cdot O_1(s_0)$ for some $w$. Then 
we obtain a branch $v_0v_1\ldots$ of $\cT$, where $v_0 =
v$ and $v_m = v_{m-1}\cdot O_1(s_m)$ for $m>0$.  
We say that $r$ is a {\em run of $\cT$ from $v$} if there exists an 
atom $X$ such that $(X,s_0) \in K(v)$,
and for each $m\geq 0$ there
exists 
a joint action $\ba \in P_e(s_m)\times P(v_m) $ such that $s_{m+1} = \tau(\ba)(s_m)$. 
That is, the actions of agent 1 labelling the branch
corresponding to $r$, together with some choice of the environment's
actions, generate the sequence of states in the run.

\newcommand{\ssat}{\models^*}

We now define a relation $\ssat$ on points of the runs from vertices
of $\cT$. 
This relation
interprets subformulae of $\psi$ by treating the
linear 
temporal operators as usual, but referring to the knowledge sets to
interpret formulae involving knowledge
or the branching time operator.  
Intuitively, $\cT, v, (r,m)
\ssat \vp$ asserts that the formula $\vp$ ``holds'' at the $m$th
vertex $v_m$ reached from $v$ along $r$, as described above. More
formally, this relation is defined by means of the following
recursion:

\be
\item $\cT, v, (r,m)\ssat p$ if $\pi_e(r(m),p)= 1$ 
\item $\cT, v, (r,m)\ssat \neg \vp $ if not $\cT, v, (r,m)\ssat  \vp $. 
\item $\cT, v, (r,m)\ssat \vp_1 \land \vp_2 $ if $\cT, v, (r,m)\ssat \vp_1 $ and $\cT, v, (r,m)\ssat \vp_2$.
\item $\cT, v, (r,m) \ssat \nxt \vp$ if $\cT, v, (r,m+1)\ssat \vp $ 
\item $\cT, v, (r,m) \ssat \vp_1 U \vp_2 $ if there exists 
$k\geq m$ such that $\cT, v, (r,l) \ssat \vp_1$ for $m\leq l<k$ and 
$\cT, v, (r,k)\ssat \vp_2$.
\item $\cT, v, (r,m) \ssat K_1\vp$ if $X(\vp) = 1$ for all $(X,s) \in
K(v_m)$, where $v_m$ is determined as above.  
\item $\cT, v, (r,m) \ssat \exists\vp$ if $X(\vp) = 1$ for some $(X,s) \in
K(v_m)$, with $r(m) = s$, where $v_m$ is determined as above.  
\ee 
We use the abbreviation $\cT,(r,m)\ssat \vp$ for $\cT, r_1(0),(r,m)
\ssat \vp$. (The choice of the vertex $r_1(0)$ here is not really
significant: it is not difficult to show that for all $k\leq m$ we
have $\cT, (r,m) \ssat \vp$ iff $\cT, r_1(k),(r[k..],m-k) \ssat \vp$.)

Define a labelled tree $\cT$ for $\psi$ and $E$ to be {\em acceptable\/}
if it satisfies the following conditions: 
\begin{description} 
\item[({\sf Real})] For all observations $o$, and for all 
$(X,s)\in K(o)$, we have $X(\psi) = 1$
and $s\in I_e$.
 
\item[({\sf Init})] For 
all 
initial states $s\in I_e$, there exists
an atom $X$ for $\psi$ such that $(X,s)$ is in $K(O_1(s))$. 

\item[({\sf Obs})] For all observations $o$ and all vertices $v$ of 
$\cT$, we have $O_1(s) = o$ for all $(X,s) \in K(v\cdot o)$. 

\item[({\sf Pred})] For all observations $o$, for all vertices $v$
other than the root, and for all 
$(X,s) \in K(v\cdot o)$,  there exists $(Y,t)\in K(v)$ and 
a joint action $\ba \in P_e(t)\times P(v)$ such that $s = \tau(\ba)(t)$.

\item[({\sf Succ})] For all vertices $v$ other than the root,
for all  $(X,s) \in K(v)$  and for all 
$\ba \in P_e(s)\times P(v)$,  if $t = \tau(\ba)(s)$ 
then there exists 
an atom $Y$ such that $(Y,t)\in K(v\cdot O_1(t))$.
\item[({\sf $\exists$sound})] For all vertices $v$, and $(X,s) \in K(v)$, 
if $X(\exists \phi) =1$ then there exists $Y$ such that $(Y,s) \in K(v)$ and $Y(\phi) =1$. 

\item[({\sf $\exists$comp})] For all vertices $v$, and $(X,s), (Y,s) \in K(v)$, 
if $X(\phi) = 1$ then $Y(\exists\phi) =1$. 

\item[({\sf Ksound})] For all vertices $v$ (other than the root) and all
$(X,s)\in K(v)$, there exists a run $r$ from $v$ such that $r(0) =
s$ and for all subformulae $\vp$ of $\psi$ we have $\cT,v,(r,0)\ssat
\vp$ iff $X(\vp) =1$.

\item[({\sf Kcomp})] For all vertices $v$ and all runs $r$ from $v$ 
there exists $(X,s)\in K(v)$ such that $r(0) = s$ and for all
subformulae $\vp$ of $\psi$ we have $\cT,v,(r,0)\ssat \vp$ iff $X(\vp)
=1$.
\end{description}

The following theorem provides the characterization of realizability
of knowledge-based specifications that forms the basis for our
synthesis procedure.

\thm 
\label{realizability}
A specification $\psi$ for a single agent is realizable in the
environment $E$ iff there exists an acceptable labelled tree for
$\psi$ in $E$.  
\ethm

\prf
We first show that if there exists an acceptable tree then the
specification is realizable. Suppose $\cT$ is an acceptable tree for
$\psi$ in $E$.  We show that the protocol $P$ for agent 1 derived from
this tree realizes $\psi$.
Let $\cI$ be the system generated by $P$ in $E$.

We claim that for all points $(r,m)$ of $\cI$ and all subformulae
$\vp$ of $\psi$ we have $\cT, (r,m) \ssat \vp$ iff $\cI,(r,m)\models
\vp$.  It follows from this that $P$ realizes $\psi$ in $E$.  For, let
$r$ be a run of $\cI$.  Take $v$ to be the vertex $r_1(0)$.  By
{\sf Init}, there exists a pair $(Y,s)\in K(v)$ with $s= r(0)$. Thus,
$r$ is a run of $\cT$ from the vertex $v$. By {\sf Kcomp}, there exists an
atom $X$ such that $(X,s)$ is in $K(v)$ and for all subformulae $\vp$
of $\psi$, we have $\cT,(r,0)\ssat \vp$ iff $X(\vp)=1$. By the claim,
we obtain in particular that $\cI,(r,0)\models \psi$ iff $X(\psi)=1$.
But by {\sf Real}, we have that $X(\psi)=1$, so $\cI,(r,0)\models \psi$
also holds. This shows that $P$ realizes $\psi$ in $E$.

The proof of the claim is by induction on the complexity of $\vp$.
The base case, when $\vp$ is an atomic proposition, is
straightforward, as are the cases where $\vp$ is built using boolean
or temporal operators from subformulae satisfying the claim. We
establish the 
cases where $\vp$ is of the form $K_1\vp'$ or $\exists \vp'$.

We first assume that $\cI, (r,m) \models K_1 \vp'$, and show
$\cT,(r,m)\ssat K_1(\vp')$. That is, for all $(X,s)\in K(r_1(m))$ we
show $X(\vp') = 1$. By {\sf Ksound}, for each $(X,s) \in K(r_1(m))$ there
exists a run $r'$ of $\cT$ from $r_1(m)$ with $r'_e(0) = s$ and $\cT,
r_1(m), r' \ssat \vp'$ iff $X(\vp')=1$. Applying {\sf Pred}
and {\sf Init}, 
this run may
be extended backwards to a run $r''$ of $\cI$ with $(r'',m)
\sim_1 (r,m)$ and $r''[m..] = r'$.  By the assumption, we have that
$\cI,(r'',m) \models \vp'$.  It follows using the induction hypothesis
that $\cT, (r'',m) \ssat \vp'$.
Note that this implies $\cT,r_1(m), r' \ssat \vp'$,
hence $X(\vp')=1$.

Conversely, we suppose that $\cT,(r,m)\ssat K_1(\vp')$ and show that
$\cI, (r,m) \models K_1 \vp'$.  Suppose that $r'$ is a run of $\cI$ 
with $(r',m)\sim_1(r,m)$. We need to prove that $\cI, (r',m) \models \vp'$.
Using {\sf Init}, $r'$ is a run of $\cT$ from $r'_1(0)$.  By {\sf Succ} and
induction, $r'[m..]$ is a run of $\cT$ from $r'_1(m)= r_1(m)$.  Thus, by {\sf Kcomp}, there
exists $(X,s)\in K(r_1(m))$ such that $r'_e(m) = s$ and for all
subformulae $\vp$ of $\psi$ we have $\cT,(r',m)\ssat \vp$ iff $X(\vp)
=1$.  By the assumption that $\cT,(r,m)\ssat K_1(\vp')$, we have that
$X(\vp')=1$ for all $(X,s)\in K(r_1(m))$.  Thus, $\cT,(r',m)\ssat
\vp'$. By the induction hypothesis it follows that $\cI,(r',m)\models
\vp'$, which is what we set out to establish. This completes the proof
of the claim, and also the proof that the existence of an
acceptable tree implies the existence of a realization.

For the case where $\vp = \exists \vp'$, we argue as follows. 
First, we assume that $\cT,(r,m)\ssat \exists(\vp')$ and show that
$\cI, (r,m) \models \exists \vp'$. From the assumption, 
there exists $X$ such that $(X,r(m)) \in K(r_1(m))$ and $X(\vp') = 1$.
By {\sf Ksound}, there exists a run $r'$ from $r_1(m)$ in $T$ such that 
$r'(0) = r(m)$ and  $\cT,r_1(m),(r',0)\ssat \exists(\vp')$. 
Let $r'' = r[0..m-1]\cdot r'$. This is a run of $P$, and we have 
$\cT,(r'',m)\ssat \vp'$. By the induction hypothesis, it follows that 
$\cI, (r'',m) \models \vp'$. Since $r''[0..m] = r[0..m]$,  it follows that 
$\cI, (r,m) \models \exists \vp'$. 

Conversely, assume that $\cI, (r,m) \models \exists \vp'$.  We show that $\cT,(r,m)\ssat \exists(\vp')$. 
From assumption, there exists a run $r'$ such that 
$r[0..m] = r'[0..m]$ and 
$\cI, (r',m) \models  \vp'$. By induction, we have $\cT,(r',m)\ssat \vp'$. 
This is equivalent to $\cT,r_1(m),  (r'[m..],0)\ssat \vp'$. By {\sf Kcomp}, 
there exists $X$ such that $(X,r(m)) \in K(r_1(m))$ and $X(\vp') = 1$. 
It follows that $\cT,(r,m)\ssat \exists(\vp')$. 
This completes the argument from the existence of an acceptable tree for $\psi$ in $E$ 
to realizability of $\psi$ in $E$. 

Next, we show that if $\psi$ is realizable in $E$ then there exists an
acceptable tree for $\psi$ and $E$. Suppose that the protocol $P$ for
agent 1 realizes $\psi$ in $E$. We construct a labelled tree $\cT$ as
follows. Let $\cI$ be the system generated by $P$ in $E$.  If $(r,m)$
is a point of $\cI$, define the atom $X(r,m)$ by $X(r,m)(\vp)=1$ iff
$\cI,(r,m)\models \vp$.  Define the function $f$ to map the point
$(r,m)$ of $\cI$ to the pair $(X(r,m),r(m))$.  For all $v$ in
$\cO^+$, define $K(v)$ to be the set of all $f(r,m)$, where $(r,m)$ is
a point of $\cI$ with $r_1(m) =v $. Define $\cT$ by 
$\cT(v) =(K(v),P(v))$ for each $v\in \cO^+$. (The label of the root can be chosen
arbitrarily.) We claim that $\cT$ is an acceptable tree for $\psi$ and $E$.

For {\sf Real}, let $v=o$ for an observation $o$, and suppose that $(X,s) \in K(v)$. 
Then there exists a run $r$ of $\cI$ such that $X=X(r,0)$ and $s=r(s)$. 
Since $\cI$ realizes $\psi$, we have that $\cI,(r,0)\models \psi$
and it is immediate that $X(\psi)=1$.  

For {\sf Init}, let $s$ be an initial state. Take $r$ to be any run of $\cI$ with $r(0) = s$
and let $X = X(r,0)$. Then $(X,s) \in K(O_1(s))$. 

For {\sf Obs}, Let $o$ be an observation and $v$ a vertex of $\cT$.
If $(X,s) \in K(v\cdot o)$, then there exists a run $r$ such that $r_1(m) = v\cdot o$, $X = X(r,m)$ 
and $s = r(m)$, where $m = |v\cdot o|$. This implies that $o = O_i(r(m)) = O_i(s)$, as required. 

For {\sf Pred}, let $o$ be an observation, $v$ a vertex of $\cT$ other than the root, and
$(X,s)\in K(v\cdot o)$. Then there exists  a run $r$ of $\cI$ such that $r_1(m) = v \cdot o$ and 
$X = X(r,m)$ and $s = r(m)$. Let $Y = X(r,m-1)$ and $t = r(m-1)$. 
Since $r_1(m) = v\cdot o$ we have $r_1(m-1) = v$. It follows that $(Y,s) \in K(v)$. 
Moreover, since $r$ is a run of $\cI$,  there exists an action $\ba\in P_e(t)\times P(v)$ such that $s = \tau(\ba)(t)$. 
This gives the conditions required for the consequent of {\sf Pred}. 

For {\sf Succ}, let $v$ be a vertex other than the root, and consider $(X,s) \in K(v)$ and 
$\ba \in \P_e(s)\times P(v)$. Let $t = \tau(\ba)(s)$. By construction, there exists a run $r$ 
such that  $r_1(m) = v$ and $X = X(r,m)$ and $s = r(m)$ , where $m = |v|$. 
Let $r'$ be any run extending $r[0..m]\cdot t$. Take $Y = X(r',m+1)$. 
Then $r'_1(m+1) = v\cdot O_i(t)$ and $r(m+1) = t$, so $(Y,t) \in K(v\cdot O_1(t))$, 
as required for {\sf Succ}. 

For {\sf $\exists$sound}, suppose that $(X,s) \in K(v)$ and $X(\exists\vp) = 1$. 
We need to show that there exists $Y$ such that $(Y,s) \in K(v)$ and $Y(\vp) = 1$.
By construction, there exists a run $r$ of $\cI$ such that $X = X(r,m)$ and $s= r(m)$, 
where $m = |v|$. Moreover, we have $\cI,(r,m) \models \exists \vp$. 
Thus, there exists a run $r'$ of $\cI$ such that $r'[0..m] = r[0..m]$ and 
$\cI,(r',m) \models \vp$. Let $Y = X(r',m)$; plainly, this satisfies 
$Y(\vp) =1$. Note that $r'(m) = r(m) = s$. 
It follows from $r'[0..m] = r[0..m]$ that $(r',m) \sim_i (r.,m)$. 
Thus, $(Y,s) \in K(v)$, and $Y$ suffices for the required conclusion. 

For {\sf $\exists$comp}, suppose $(X,s),(Y,s) \in K(v)$ and $X(\phi) = 1$. 
We show that $Y(\exists\vp)=1$. By construction, there exist runs 
$r,r'$ such that $X=X(r,m)$ and $Y= Y(r',m)$ and $r_1(m) = r'_i(m) = v$
and $r(m) = r'(m) = s$. Since $X(\vp) = 1$ we have $\cI,(r,m) \models \vp$. 
Consider the sequence $r'' = r[0..m-1]r'[m..]$. This is a run of $\cI$ with 
$r''[0..m] = r[0..m]$. Thus, $\cI,(r'',m) \models \exists \vp$. 
Since satisfaction of formulas depends only on the 
future
and $r''[m..] = r'[m..]$, we obtain that $\cI,(r',m) \models \exists \vp$. 
This yields that $Y(\exists \vp) = 1$, as required.

We next prove {\sf Ksound} and {\sf Kcomp}.
For this, we first prove that for all points $(r,m)$ of $\cI$ we have
$\cI,(r,m)\models \vp$ iff $\cT,(r,m)\ssat \vp$.  The proof is by
induction on the complexity of $\vp$.  As above, the cases not
involving knowledge are straightforward, so we focus on the case where
$\vp$ is of the form $K_1\vp'$. By definition, $\cI,(r,m)\models
K_1\vp'$ iff $\cI,(r',m)\models \vp'$ for all $(r',m)\sim_1 (r,m)$. By
definition of $\cT$ and the induction hypothesis, this holds just when
$X(\vp')=1$ for all $(X,s)\in K(r_1(m))$.  This latter condition is
equivalent to $\cT,(r,m)\ssat K_1\vp'$, so we are done.

For {\sf Ksound}, suppose that $v$ is a vertex not equal to the root and
that $(X,s)$ is in $K(v)$.  Then there exists a point $(r,m)$ of $\cI$
such that $v = r_1(m)$ and $(X,s)=f(r,m)$.  The sequence $r[m..]$ is a
run of $\cT$ from $v$ with initial state $s$. To establish {\sf Ksound},
we need to show that for all subformulae $\vp$ of $\psi$ we have
$X(\vp)=1$ iff $\cT,v,r[m..]\ssat \vp$.  This holds because
$\cI,(r,m)\models \vp$ iff $\cT,(r,m)\ssat \vp$.

We now prove that $\cT$ satisfies {\sf Kcomp}. Let $v$ be a vertex of
$\cT$ not equal to the root and let $r$ be a run of $\cT$ from $v$.
We need to show that there exists a pair $(X,s)$ in $K(v)$ such that
$s=r(0)$ and, for all subformulae $\vp$ of $\psi$, $X(\vp)=1$ iff
$\cT,v,r\ssat \vp$.  By definition of a run from $v$, there exists
$(Y,s)\in K(v)$ with $r(0) = s$. By construction of $\cT$, there
exists a point $(r',m)$ of $\cI$ such that $(Y,s) = f(r',m)$.
Clearly, the sequence $r'' = r'[0..m-1]\cdot r$ is a run of
$\cI$. Thus, we have that $f(r'',m)= (X(r'',m),s)$ is in $K(v)$.  By
definition we have $X(r'',m)(\vp)=1$ iff $\cI,(r'',m)\models \vp$. As
shown above, the latter holds just when $\cT,(r'',m)\ssat \vp$.  But
this last condition is equivalent to $\cT,v,r\ssat \vp$, since
$r''[m..] = r$. This shows that $f(r'',m)$ is the required pair
$(X,s)$. 
\eprf

In the next section, we show how this result can be used to yield an
automata-theoretic procedure for constructing a realization of a
specification.

\section{An Algorithm for Realizability} \label{sec:alg} 
We first recall the definitions of the two types of automata we require. 
Section~\ref{sec:aut-words}  deals with automata on infinite
words, and Section~\ref{sec:aut-tree} deals with alternating automata on 
infinite trees. We apply these to our realizability problem in Section~\ref{sec:realize}.

\subsection{Automata on Infinite Words} \label{sec:aut-words} 

For an introduction to the theory of automata on infinite words and
trees see \cite{Tho90}.

The types of finite automata on infinite words we consider
are those defined by B\"uchi \cite{Buc62}.
A (nondeterministic) automaton on words is a tuple
$\cA=\zug{\Sigma,S,S_0,\rho,\alpha}$, where
$\Sigma $ is a finite alphabet,
$S$ is a finite set of states,
$S_0 \subseteq S$ is a set of starting states,
$ \rho : S \times \Sigma \rightarrow 2^S $ is a
(nondeterministic) transition function, and
$\alpha$ is an acceptance condition.
A B\"{u}chi acceptance condition is a set $F\subseteq S$.

A {\em run\/} $r$ of $\cA$ over a infinite word $w=a_0 a_1 \cdots$, is a
sequence $s_0,s_1 ,\cdots$, where $s_0 \in S_0$ and
$s_i \in \rho (s_{i-1} , a_{i-1})$, for all $i \ge 1$.
Let ${\it inf}(r)$ denote the set
of states in $Q$ that appear in $r(\rho)$ infinitely often.
The run $r$ satisfies a B\"{u}chi condition $F$
if there is some state in $F$ that repeats infinitely often in $r$,
i.e., $F \cap {\it inf}(r) \not=\emptyset$.
The run $r$ is {\em accepting\/} if it satisfies the acceptance
condition, and the infinite word $w$ is {\em accepted\/}
by $\cA$ if there is an accepting run of $A$ over $w$.
The set of infinite words accepted by $\cA$ is denoted
$\cL(\cA)$.
 
The following theorem establishes the correspondence between temporal
formulae and B\"uchi automata.
\pro\label{VW}
{\rm \cite{VW94}}
Given a temporal formula $\phi$ over a set $Prop$ of
propositions, one can build a B\"uchi automaton
$\cA_\phi =\zug{2^{Prop},S,S_0,\rho,F} $, where
${\vert}S{\vert} \le 2^{O({\vert}\phi{\vert})}$,
such that $\cL(\cA_\phi) $ is exactly the set of computations
satisfying the formula $\phi$.
\epro

\subsection{Alternating Automata on Infinite Trees}\label{sec:aut-tree}

{\em Alternating tree automata} generalize
nondeterministic tree automata and were first introduced in \cite{MS87}.
They have recently found usage in computer-aided verification
\cite{BVW94,Var95d,Var97}.
An alternating tree automaton
${\cal A}=\zug{\Sigma,Q,q_0,\delta,\alpha}$ runs on
$\Sigma$-labelled $\Upsilon$-trees 
(i.e., mappings from $\Upsilon^*$ to $\Sigma$). 
It consists of a finite set $Q$ of states, an initial 
state $q_{0} \in Q$, a transition function $\delta$, and an acceptance
condition $\alpha$ (a condition that defines a subset of $Q^\omega$).
 
For a set $D$, let 
${\cal B}^{+}(D)$ 
be the set of positive Boolean formulae over $D$;
i.e., Boolean formulae built from elements in $D$ using
$\wedge$ and $\vee$, where we also allow the formulae $\bft$ and
$\bff$.  
For a set $C \subseteq D$ and a formula $\theta \in {\cal B}^{+}(D)$,
we say that $C$ {\em satisfies} $\theta$ iff assigning $\bft$ to
elements in $C$ and assigning $\bff$ to elements in $D \setminus C$
makes $\theta$ true.

The transition function 
$\delta : Q \times \Sigma \rightarrow {\cal B}^{+}(\Upsilon \times Q)$ 
maps a state and an input letter to a formula that
suggests a new configuration for the automaton.
A {\em run\/} of an alternating automaton ${\cal A}$ on an input
$\Sigma$-labelled $\Upsilon$-tree $\cT$ is a tree $\zug{T_r,r}$
in which the root is labelled by $q_0$ and every other node is 
labelled by an element of $\Upsilon^* \times Q$.
Here $T_r$ is a prefix-closed subset of $\boldN^*$ and $r:T_r\rightarrow \Upsilon^* \times Q$ is the labeling function.
Each node of $T_r$ corresponds to a node of $\Upsilon^*$.
A node  
$y$ in $T_r$, labelled by $r(y)=(x,q)$, describes a copy of the automaton
that reads the node $x$ of $\Upsilon^*$ and visits the state $q$.
Formally, $\zug{T_r,r}$ is a $\Sigma_r$-labeled tree where
$\Sigma_r=\Upsilon^* \times Q$ and $\zug{T_r,r}$
satisfies the following:
\begin{enumerate}
\item
$\epsilon \in T_r$ and $r(\epsilon)=(\epsilon,q_0)$.
\item
Let $y \in T_r$ with $r(y)=(x,q)$ and $\delta(q,\cT(x))=\theta$.
Then there is a (possibly empty) set
$S=\{(c_1,q_1),\ldots,(c_n,q_n)\} \subseteq
\Upsilon \times Q$, such that the following hold:
\begin{itemize}
\item
$S$ satisfies $\theta$, and
\item
for all $1 \leq i \leq n$, we have $y \cdot i \in T_r$ and
$r(y \cdot i)=(x \cdot c_i,q_i)$.
\end{itemize}
\end{enumerate}
For example, if $\zug{T,V}$ is a $\{0,1\}$-tree with $V(\epsilon)=a$ 
and 
$\delta(q_0,a)=((0,q_1) \vee (0,q_2)) \wedge ((0,q_3) \vee (1,q_2))$,
then the nodes of $\zug{T_r,r}$ at level $1$ include the label
$(0,q_1)$ or $(0,q_2)$, and include the label $(0,q_3)$ or $(1,q_2)$.

Each infinite path $\rho$ in  $\zug{T_r,r}$ is labelled by
a word $r(\rho)$ in $Q^\omega$. 
A run $\zug{T_r,r}$ is accepting iff all its infinite paths satisfy
the acceptance condition.
Let ${\it inf}(\rho)$ denote the set
of states in $Q$ that appear in $r(\rho)$ infinitely often.
In a B\"uchi acceptance condition, $\alpha \subseteq Q$
and an infinite path $\rho$ satisfies an acceptance condition
$\alpha$ if $\alpha \cap {\it inf}(\rho) \neq\emptyset$,
In a co-B\"uchi acceptance condition, $\alpha \subseteq Q$
and an infinite path $\rho$ satisfies an acceptance condition
$\alpha$ if $\alpha \cap {\it inf}(\rho) =\emptyset$,
In a {\em Rabin} acceptance condition, $\alpha \subseteq 2^Q \times
2^Q$, and an infinite path $\rho$ satisfies an acceptance condition
$\alpha= \{\zug{G_1,B_1},\ldots,\zug{G_m,B_m}\}$ iff
there exists $1 \leq i \leq m$ for which ${\it inf}(\rho) \cap G_i
\neq \emptyset$ and ${\it inf}(\rho) \cap B_i = \emptyset$.
As with nondeterministic automata, an automaton accepts a tree iff
there exists an accepting run on it.
We denote by ${\cal L}({\cal A})$ the language of the automaton
${\cal A}$; i.e., the set of all labelled trees that ${\cal A}$
accepts. 
${\cal A}$ is {\em empty\/} if ${\cal L}({\cal A})=\emptyset$.

{\em Nondeterministic\/} tree automata are a special case of 
alternating tree automata.
An automaton ${\cal A}=\zug{\Sigma,Q,q_0,\delta,\alpha}$ is
nondeterministic if, for each state $q\in Q$ and letter
$a\in\Sigma$, the formula $\delta(q,a)$ does not contain
two pairs $(c,q_1)$ and $(c,q_2)$, where $q_1 \neq q_2$,
that are conjunctively related (i.e., both appear in the same
disjunct of the disjunctive normal form of $\delta(q,a)$).
Intuitively, it means that the automaton cannot send two
distinct copies in the same direction \cite{MS87}.

\pro\label{MS}
{\rm \cite{MS95}}
Given an alternating Rabin automaton with $n$ states and $m$ pairs,
we can translate it into an equivalent nondeterministic Rabin automaton
with $(mn)^{O(mn)}$ states and $mn$ pairs.
\epro

\pro\label{EJPR}
{\rm \cite{EJ88,PR89a,KV98}}
Emptiness of a nondeterministic Rabin automaton with $n$ states and 
$m$ pairs over an alphabet with $l$ letters can be tested in time
$(lmn)^{O(m)}$.
\epro

\subsection{Realizability} \label{sec:realize} 

We now derive an automata-theoretic algorithm for realizability for
knowledge-based specifications involving a single agent.

\thm\label{psitoA}
There is an algorithm that constructs for a given
specification $\psi$ and an environment $E$ an 
nondeterministic
Rabin automaton $\cA_{\psi,E}$ such that $\cA_{\psi,E}$ accepts 
precisely the acceptable trees for $\psi$ in $E$.
The automaton 
$\cA_{\psi,E}$ has $2^{||E||\cdot 2^{O(||\psi||)}}$ states
and $||E||\cdot 2^{O(||\psi||)}$ pairs.
\ethm
\prf
(sketch)
The inputs to the automaton $\cA_{\psi,E}$ are
$\labL$-labeled trees. Note that the size of $\labL$
is exponential in the number of states 
and actions 
in $E$ and doubly
exponential in the length of $\psi$.

To check that an input tree $\cT$ is acceptable, the automaton
has to check that it satisfies the properties {\sf Real},
{\sf Init}, {\sf Obs}, {\sf Pred}, {\sf Succ}, 
{\sf $\exists$sound}, {\sf $\exists$comp}, 
{\sf Ksound},
and {\sf Kcomp}.
We describe automata that check these properties;
$\cA_{\psi,E}$ is obtained as the intersection of these automata.
The property {\sf Real} is a condition on the
children of the root of $\cT$ that can be checked by a
nondeterministic automaton with $O(1)$ states.
The property {\sf Init} is a condition on the
children of the root of $\cT$ that can be checked by a
nondeterministic automaton with $O(||E||)$ states.
The property {\sf Obs} is a condition on the labels of nodes
in the tree that can be checked by a nondeterministic
automaton with $O(||E||)$ states.

The property {\sf Pred} is a condition on relationships between
labels of nodes and labels of their children that can be checked
by a nondeterministic
automaton with $2^{O(||E||)}$ states; after visiting a node $v$, 
the automaton remembers $P(v)$ and the set $\{t : (Y,t) \in K(v)\}$.
It then uses this to check that for all $(X,s)\in K(v\cdot o)$
there exists $(Y,t)\in K(v)$ and 
$\ba\in P_e(t)\times P(v)$ such that $s=\tau(\ba)(t)$.
The property {\sf Succ} is a condition on relationships between
labels of nodes and labels of their children that can be checked
by a nondeterministic
automaton with $2^{O(||E||)}$ states; after visiting a node $v$, 
the automaton remembers $P(v)$ and the set $\{t : (Y,t) \in K(v)\}$.
It then uses this to check that for all $(Y,s)\in K(v)$
and 
$\ba\in P_e(s)\times P(v)$ there exists $(Y,t)\in K(v\cdot o)$ such that $t=\tau(\ba)(s)$.

Conditions {\sf $\exists$sound} and {\sf $\exists$comp} are 
trivially checked conditions on the labels. 

To check {\sf Ksound}, an alternating automaton guesses,
for all vertices $v$ (other than the root) and all
$(X,s)\in K(v)$, a run $r$ from $v$ such that $r(0)=s$ 
and for all subformulae $\vp$ of $\psi$ we have 
$\cT,v,(r,0)\ssat \vp$ iff $X(\vp) =1$.
A formula $\xi$ can be viewed as a temporal formula by
considering every subformula $K\theta$ 
or $\exists\theta$
as a new proposition.
Consider the formula $\psi_X$ that is obtained by taking
the conjunction of subformulae of $\psi$ or their negation
according to $X$. We consider $\psi_X$ as a temporal formula
in LTL 
and appeal to Theorem~\ref{VW} to construct
a B\"uchi automaton $A_{\psi_X}$ that check whether $\psi_X$ is
satisfied by 
a
sequence of truth assignments to its extended set
of propositions (i.e., atomic propositions and subformulae
of the form $K\theta$).
Thus, the automaton guesses a sequence $v_0,v_1,\ldots$
of nodes in the tree and a sequence $(X_0,s_0), (X_1,s_1),\ldots$
of atom-state pairs such that $v_0=v$, $X_0=X$, $s_0=s$,
$v_{i+1}$ is a child of $v_i$, $(X_i,s_i)\in K(v_i)$, and
$s_{i+1}=\tau(\ba)(s_i)$ for some $\ba\in P_e(s_i)\times P(v_i)$.
It then emulates $A_{\psi_X}$ and checks that 
the sequence $X_0,X_1,\ldots$ is accepted.
This automaton has $||E||\cdot 2^{O(||\psi||)}$ states and
a B\"uchi acceptance condition.

Instead of checking that {\sf Kcomp} holds, we construct
an alternating automaton that checks that {\sf Kcomp} is violated, 
since alternating automata can be complemented by dualizing their
transition function (i.e., switching $\vee$ and $\wedge$ as well
as $\bft$ and $\bff$) and complementing the acceptance condition
\cite{MS87}.
The automaton guesses a vertex $v$ and a run $r$ from $v$ such
that for no $(X,s)\in K(v)$ we have that $r(0) = s$ and for all
subformulae $\vp$ of $\psi$ we have $\cT,v,(r,0)\ssat \vp$ iff 
$X(\vp)=1$.
We already saw how the automaton guesses a run;
it guesses a sequence $v_0,v_1,\ldots$
of nodes in the tree and a sequence $(X_0,s_0), (X_1,s_1),\ldots$
of atom-state pairs such that $v_0=v$, $(Y_0,s_0)\in K(v)$
for some atom $Y_0$, but $(X_0,s_0) \not\in K(v)$,
$v_{i+1}$ is a child of $v_i$, $(Y_i,s_i)\in K(v_i)$
for some atom $Y_i$, and
$s_{i+1}=\tau(\ba)(s_i)$ for some $\ba\in P_e(s_i)\times P(v_i)$.
It then emulates $A_{\psi_{X_0}}$ and checks that 
the sequence $X_0,X_1,\ldots$ is accepted.
This automaton has $||E||\cdot 2^{O(||\psi||)}$ states.
After complementing it, it has a co-B\"uchi acceptance
condition.

We now apply Proposition~\ref{MS} to the alternating automata
that check {\sf Ksound} and {\sf Kcomp} to get nondeterministic
Rabin automata with $2^{||E||\cdot 2^{O(||\psi||)}}$ states
and $||E||\cdot 2^{O(||\psi||)}$ pairs.
\eprf

\cor\label{complexity}
There is an algorithm that decides whether a formula $\psi$ is
realizable in an environment $E$ in time 
$2^{O(||E||)}\cdot 
2^{2^{O(||\psi||)}}$.
\ecor
\prf
By Theorem~\ref{psitoA}, $\psi$ is realizable in $E$ iff
$\cL({\cal A}_{\psi,E})\neq\emptyset$.
The claim now follows by Proposition~\ref{EJPR}, 
since ${\cal A}_{\psi,E}$ has
Rabin automata with $2^{||E||\cdot 2^{O(||\psi||)}}$ states
and $||E||\cdot 2^{O(||\psi||)}$ pairs and the alphabet has
$2^{||E||\cdot 2^{O(||\psi||)}}$ letters.
\eprf

\noindent
We note that it is shown in \cite{PR89a} that realizability of
temporal formulae with complete information is already 2EXPTIME-hard.
Thus, the bound in Corollary~\ref{complexity} is essentially
optimal.

So far our focus was on realizability.
Recall, however, that if $\cT$ is an acceptable tree for
$\psi$ in $E$, then the protocol $P$ for agent 1 derived from
this tree realizes $\psi$ in $E$.
The emptiness-testing algorithm used in the realizability test
(per Proposition~\ref{EJPR}) does more than just test emptiness.
When the automaton is nonempty the algorithm returns a 
{\em finitely-generated\/} tree, which, as shown in \cite{BL69},
can be viewed as a finite-state protocol.
We return to this point in the following section.

\section{Discussion} \label{sec:discn}

In this section we make a number of remarks concerning realizability
of specifications involving knowledge. We first consider, in
section~\ref{sec:kii}, the question of what a protocol realizing a
specification knows. Then, in section~\ref{sec:kbp}, we relate 
realizability of specifications involving knowledge to 
\emph{knowledge-based programs}. 

\subsection{Knowledge in the Implementation} \label{sec:kii}

In this section we remark upon a subtle point concerning the states of
knowledge attained in protocols realizing a specification. As these
remarks apply equally to the general multi-agent framework we have
defined, we return to this context.

We have defined local states, hence the semantics of knowledge, using
the assumption of synchronous perfect recall, which involves an
infinite space of local states.  A protocol realizing a specification
is not required to have perfect recall, and could well be represented
(like the protocol synthesized by our procedure) using a finite set of
states. The sense in which such a protocol satisfies the conditions on
knowledge stated by the specification is the following: an agent that
follows the actions prescribed by the protocol, but computes its
knowledge based on the full record of its observations, satisfies this
specification. Thus, although we may have a finite-state protocol, it
appears that we have not in actuality eliminated the need to maintain
an unbounded log of all the agent's observations. If this is so, then
the system is better characterized as consisting of an infinite state
space coupled to a finite-state controller.

Now, there are situations in which we can dispense with the observation
logs, leaving just the finite-state controller. This holds when,
although we state the specification in knowledge-theoretic terms, we
are more concerned with the \emph{behavior} of the synthesized system
than the information encoded in its states. For example, Halpern and
Zuck \cite{HZ} give a knowledge-based specification (in the form of a
knowledge-based program) of solutions to a sequence transmission
problem. They start with the assumption of perfect recall, but their
ultimate interest is to develop implementations for this specification
that optimize the memory maintained by agents while preserving their
behaviour. One of the implementations they consider, the 
alternating-bit protocol \cite{BSW}, is a finite-state protocol.  
(We remark that the
definitions of implementation of knowledge-based programs in
\cite{FHMVdc97,FHMV} do not admit such optimized protocols as
implementations of knowledge-based programs, but the modified approach
of \cite{meyden:tark96} does.)

One might wonder whether, if we only wish to specify behaviour, one 
can state an equivalent specification that makes no use of the
knowledge operators.  This is not the case: the knowledge operators
add expressive power, making it possible to specify that the behavior
is \emph{information-theoretically optimal}. For example, the 
knowledge operators allow one to specify that the agent performs an
action \emph{as soon as} it has information appropriate to that
action. A simple example of such a specification is studied by Brafman
et al.\ \cite{BrafmanLMS1997}, who consider a robot provided with
incomplete information about its location through a noisy position
sensor. The robot must satisfy the specification that it halts as soon
as it knows that it is inside a goal region. Other examples
of behavioral specifications with information-theoretic optimality
constraints have been considered in a sequence of papers on agreement
protocols in distributed systems \cite{HM90,DM,MT,NeigerBazzi}.

Although in some cases one is concerned only with behavior, in others
what one has in mind in writing a knowledge-based specification is to
construct an implementation whose states have the 
information-theoretic property expressed. 
This is the case when the states of 
knowledge in question function as an output of the system, or provide 
inputs to some larger module. For example, we might specify that
a controller for a nuclear reactor must keep the reactor temperature 
below a certain level and must also know of a critical 
level of radiation whenever this condition holds, with the intention 
that this information be provided to the operator. In this case
it will not do to implement the specification  according to its 
behavioral component alone, since this might lose the attribute, 
knowledge of radioactivity, that we wish to present as an output. 

Clearly, we could always ensure that the knowledge properties
specified are available in the implementation by taking the
implementation to consist of both the finite-state controller and the
log of all the agent's observations.  Such an implementation is rather
inefficient.  Can we do better?  One attempt to do so would be simply
to take the implementation to consist just of the protocol, and to
compute knowledge on the basis of the protocol states.

To make this idea precise, we adopt the following model of a
protocol and the knowledge it encodes. We suppose that 
agent $i$'s protocol is represented as an automaton $A_i = \la
Q_i, q_i,\mu_i, \alpha_i\ra$, where
\begin{enumerate} 
\item $Q_i$ is the  set of {\em protocol states}, 
\item $q_i\in Q$ is the initial state, 
\item $\mu_i:Q_i\times \cO \rightarrow Q_i$ is the state transition function, 
used  to update the protocol state given an observation in $\cO$, and 
\item 
$\alpha_i: Q_i\rightarrow \pow{\act_i}$ is a function mapping each state to a set of actions of the agent. 
\end{enumerate} 
As usual, we define the state reached after a sequence $\sigma$ of
inputs (i.e., observations of the agent) by $A_i(\epsilon) = q_i$ and
$A_i(\sigma \cdot o) = \mu_i(A_i(\sigma),o)$.  We may then define the
protocol itself, as a function from sequences of observations to
sets of 
actions, by $P_{A_i}(\sigma) =
\alpha_i(A_i(\sigma))$. 

\newcommand{\peqi}{\approx_i^A}

Suppose we are given a tuple $A = \la A_1, \ldots, A_n\ra$
of automata representing the protocols of agents $1\ldots n$. 
To interpret the knowledge operators with respect to the states of
these automata, we first define, for each agent $i$, 
an indistinguishability relation
$\peqi$ on points, based on the states of the automata $A_i$ rather than
the perfect-recall local states used for the relation 
$\eqi$ above. That is, we define 
$(r,m) \peqi (r',m')$ to hold when $A_i(r_i(m)) = A_i(r'_i(m'))$. 
We may now define the semantics of knowledge exactly as we did using 
the relation $\eqi$. 
To distinguish the two interpretations, we 
introduce new knowledge modalities $K_i^A$, and define 
$\cI,(r,m) \models K_i^A\vp$ if $\cI,(r',m') \models \vp$
for all points $(r',m')$ of $\cI$ satisfying $(r',m')\peqi(r,m)$. 
We may now formulate the proposal above as follows. 
Suppose a specification $\vp$ is realized in an environment $E$ by a
joint protocol $\bP_A$, represented by the automata $A$. 
Is it then the case that this joint protocol realizes in $E$ the specification
$\vp^A$ 
obtained from $\vp$ by replacing (recursively) each subformula
$K_i \psi$ with $K_i^A\psi$? 
It is not, as the following example shows.
\begin{example} 
The protocol in Example~\ref{ex:tog}, which performs the toggle action
at all steps, can be represented by an automaton $A$ with a single
state. This protocol realizes the specification $\Box ( K_1
\mbox{\tt toggle-on} \Or K_1 \neg \mbox{\tt toggle-on})$. However, with respect to the
automaton $A$, the formula $\Box ( K_1^A \mbox{\tt toggle-on} \Or K_1^A \neg
\mbox{\tt toggle-on})$ is false at time 0 in a run generated by the
protocol. For, at even numbered points on these runs the toggle is on and at
odd points the toggle is off, and the single state does not suffice to
distinguish the two. $\Box$
\end{example}

Nevertheless,
a slight modification of the proposal makes it possible to ensure that
the protocols realizing a specification have the desired information
theoretic property.
All that is required is to reflect an agent's knowledge according 
to the perfect-recall definition in its behavior.
To do so, we first modify the environment so that an 
agent is provided with actions that allow it to assert what it knows, 
and then add a constraint to the specification that requires agents 
to assert their knowledge truthfully.

Suppose that $\Phi$ is the set of all the knowledge formulae of the
form $K_i(\vp)$, for some $i=i..n$,  
that we wish the implementation to preserve, together
with all subformulae of such formulae that have the same form. 
For each $i=1..n$, let $\Phi_i$ be the set of formulas in $\Phi$ of the form $K_i\vp$. 
The
modification of the environment involves adding to each 
agent's
actions a component in which the agent asserts a subset of
$\Phi_i$. That is, we take $\act_i'$ to be $\act_i \times {\cal
P}(\Phi_i)$. We also modify the environment $E= \la S_e, I_e, P_e,
\tau , O_1,\ldots, O_n, ,\pi_e \ra$ to the environment $E'= \la S_e',
I_e', P_e', \tau' , O_1',\ldots O_n',\pi_e' \ra$, where 
\be 
\item $S_e'= S_e \times {\cal P}(\Phi_1)\times \ldots \times {\cal P}(\Phi_n)$; intuitively, a state 
$(s,\Psi_1, \ldots , \Psi_n)\in S_e'$ represents that the state of the environment is $s$ and each 
agent $i$'s last assertion is $\Psi_i$,
\item $I_e' = I_e \times \{\emptyset\}\times \ldots \{\emptyset\}$; we take the last assertion to be 
empty at the initial state,
\item $P_e'(s,\Psi_1, \ldots, \Psi_n) = P_e(s)$ for all $s\in S_e$ and $\Psi_1, \ldots, \Psi_n \subseteq \Phi$; 
so that the latest assertion has no impact on the environment's protocol, 
\item $\tau'(\la a_e, (a_1,\Psi'_1), \ldots, (a_n, \Psi_n') \ra)[(s,\Psi_1, \ldots, \Psi_n)] = (\tau(\la a_e, a_1, \ldots a_n \ra )(s), 
\Psi'_1, \ldots, \Psi'_n)$, so that state transitions operate as in $E$, but additionally record 
each agent's latest assertion, 
\item $O_i'(s,\Psi_1, \ldots , \Psi_n) = O_i(s)$, so that the agent's observations are 
unchanged, and 
\item $\pi_e'((s,\Psi_1, \ldots, \Psi_n),p) = \pi_e(s,p)$ for all atomic propositions $p\in \Prop$. 
\ee 
Additionally, we extend the language by introducing for each formula
$\psi \in \Phi_i$ an atomic proposition ``$\said_i(\psi)$'', with semantics
given by $\pi_e'((s,\Psi_1, \ldots, \Psi_n),\said_i(\psi)) = 1$ iff $\psi \in \Psi_i$.

Define ${\rm Say}(\Phi)$ to be the formula 
\[ 
\bigwedge_{i=1..n}~
\bigwedge_{K_i\psi\in \Phi} \Box (K_i\psi \equiv 
\nxt \ \said_i(K_i\psi)), 
\] 
which asserts that agents say what they know (according to 
perfect recall.) Additionally, define 
${\rm Know}(\Phi,A)$ 
to be the formula 
\[ 
\bigwedge_{i=1..n}~
\bigwedge_{K_i\psi\in \Phi} \Box (K_i\psi \equiv 
K_i^A\psi)), 
\] 
which says that each agent knows a fact in $\Phi$ according 
to its protocol just when it knows this fact using perfect 
recall. 
We then have the following result.

\pro
The following are equivalent: 
\be 
\item 
For some (finite state) automaton $A$, the formula $\vp \And {\rm Know}(\Phi,A)$ is realized in $\cI(P_A, E)$. 
\item 
The formula $\vp \And {\rm Say}(\Phi)$ is (finite state) realizable in $E'$. 
\ee
\epro

\prf 
Suppose first that  $\vp \And {\rm Know}(\Phi,A)$ is realized in $E$, by the joint protocol represented
by the (finite state) automata $A= \la A_1, \ldots, ,A_n\ra$. We construct automata $A'= \la A'_1, \ldots, ,A'_n\ra$
for the environment $E'$ such that the corresponding joint protocol realizes $\vp \And {\rm Say}(\Phi)$.  
For each  $A_i = \la Q_i, q_i,\mu_i, \alpha_i\ra$, define $A'_i$ to be identical to $A_i$ except that it 
has action function $\alpha_i'$, defined by 
$\alpha'_i(q) = \alpha_i(q)\times\{ \Phi_q\}$ where 
$$\Phi_q = 
\{K_i\psi\in \Phi~|~ \cI(P_A,E), (r,m) \models K^A_i\psi~\mbox{for some $(r,m)$ with $A_i(r_i(m)) = q$}\}$$
for all $q\in Q_i$.
Plainly, $A'_i$ is finite state if $A_i$ is. Observe, moreover, that 
$\cI(P_A,E), (r,m) \models K^A_i\psi$ for some $(r,m)$ with $A_i(r_i(m)) = q$ iff 
$\cI(P_A,E), (r,m) \models K^A_i\psi$ for all $(r,m)$ with $A_i(r_i(m)) = q$. 

We note that the runs of $\cI(P_A,E)$ and $\cI(P_{A'},E')$ are in one-to-one correspondence, with 
a run $r'$ of the latter with $r'(m) = (s,\Psi_1, \ldots, \Psi_n)$ corresponding to a run $r$ of the former 
with $r(m) = s$. It is immediate from this and the fact that  $\cI(P_A,E)$ realizes $\vp$  that 
$\cI(P_{A'},E')$ realizes $\vp$. It therefore remains to show that $\cI(P_{A'},E')$ realizes ${\rm Say}(\Phi)$. 
Let $K_i\psi \in \Phi$ and let $r'$ be a run of $\cI(P_{A'},E')$ corresponding to run  $r$ of $\cI(P_A,E)$. 
Then, by definition of $A_i'$, and the observation above, 
we have $\cI(P_{A'},E'),(r',m) \models \nxt\said_i(K_i\psi)$ iff 
$\cI(P_{A},E),(r,m) \models K^A_i\psi$. 
By the fact that $\vp \And {\rm Know}(\Phi,A)$ is realized in $E$, 
we have that $\cI(P_{A},E),(r,m) \models K^A_i\psi$ iff $\cI(P_{A},E),(r,m) \models K_i\psi$. 
By the correspondence noted, we have 
$\cI(P_{A},E),(r,m) \models K_i\psi$ iff $\cI(P_{A'},E'),(r',m) \models K_i\psi$. 
It follows that $\cI(P_{A'},E'),(r',m) \models K_i\psi \dimp \nxt\said_i(K_i\psi)$, as required for 
$\cI(P_{A'},E') \models {\rm Say}(\Phi)$. 

Conversely, suppose first that  $\vp \And {\rm Say}(\Phi)$ is realized in $E'$, by the joint protocol represented
by the (finite state) automata $A'= \la A'_1, \ldots, ,A'_n\ra$. We construct automata $A= \la A_1, \ldots, ,A_n\ra$
for the environment $E$ such that the corresponding joint protocol realizes $\vp \And {\rm Know}(\Phi,A)$.  
For each $i=1..n$, the automaton $A_i$ is obtained from $A_i'$ by replacing the action function $\alpha'_i$ 
by the function $\alpha_i$, such that 
$a \in \alpha_i(q) = a$ iff there exists $\Psi$ such that $(a,\Psi) \in \alpha'(q)$. 
Since the state sets are the same, $A_i$ is finite state iff $A'_i$ is. 

Note first that for \emph{every} automaton $A$, we have that if
$(r,m)\eqi(r',m')$ then $(r,m)\peqi(r',m')$, since the automaton is
fed the same sequence of inputs in reaching the point $(r,m)$ and
$(r',m')$.  It follows that every automaton 
$A$ realizes the formula
$\bigwedge_{i=1..n}~ \bigwedge_{K_i\psi\in \Phi} \Box (K_i^A\psi \rimp K_i\psi))$. 
Thus, we need to establish the converse. 

Note that it follows from the fact that $\cI(P_{A'}, E')$ realizes $ {\rm Say}(\Phi)$ that 
for all points $(r,m)$ of $\cI(P_{A'}, E')$, if $A'_i(r_i(m))= q$ and $(a,\Psi), (a',\Psi')\in \alpha_i'(q)$,  then 
$\Psi = \Psi'$. It follows, by construction of $A, A'$ and $E'$, that the runs of 
$\cI(P_A,E)$ and $\cI(P_{A'},E')$ are in one to one correspondence,
the only difference being that in the latter the agents additionally 
assert some set of formulae. This does not affect satisfaction of 
formulae in the language based on $\Prop$ in any way. 
Suppose that $\cI(P_A,E),(r,m) \models K_i\psi$, where $K_i\psi\in
\Phi$. We show that 
$\cI(P_A,E),(r,m) \models K_i^A\psi$. Let $(\rho,k)$ be a point of 
$\cI(P_A,E)$ such that $(\rho,k) \peqi(r,m)$, i.e., 
$A(\rho_i(k)) = A(r_i(m))$. Since the states and transition functions of $A$ and $A'$ are 
identical, we also have $A'(\rho_i(k)) = A'(r_i(m))$. 

Let $r',\rho'$ be the runs of $\cI(P_{A'},E')$ corresponding to $r,\rho$, respectively, 
and let $ (a,\Psi)= A'(r'_i(m))$.  Since $\cI(P_A,E),(r,m) \models K_i\psi$, we have that 
$\cI(P_{A'},E'),(r',m) \models K_i\psi$. 
Since $\cI(P_{A'},E')$ realizes ${\rm Say}(\Phi)$, this implies that $K_i\psi \in \Psi$.  
Since $A'(\rho_i(k)) = A'(r_i(m))$, we also have $  A'(\rho_i(k))= (a,\Psi)$
and $K_i\psi \in \Psi$.  Because $\cI(P_{A'},E')$ realizes ${\rm Say}(\Phi)$, this implies that 
$\cI(P_{A'},E'),(\rho',k) \models K_i\psi$. By the correspondence, it follows that 
$\cI(P_A,E),(\rho,k) \models K_i\psi$, from which we obtain that 
$\cI(P_A,E),(\rho,k) \models \psi$. Since this holds for all $(\rho,k) \peqi (r,m)$, 
we have $\cI(P_A,E),(r,m) \models K^A_i\psi$. Thus, we have shown that
$\cI(P_{A},E),(r,m)\models K_i^A\psi \rimp K_i \psi$.
\eprf

Intuitively, this result holds because the implementation can only
behave as specified by $\vp \And {\rm Say(\Phi)}$ if the protocol
states encode the relevant knowledge.  This result shows that,
provided some care is taken in writing specifications, the
realizability framework we have defined in this paper is capable of
handling both the case in which agents are required simply to 
behave as if they had perfect recall, and the case in which agents  are
required both to behave in this fashion \emph{and} encode certain 
perfect-recall knowledge in their protocol states. 
(Note that it follows from the fact that $\vp \And {\rm Know}(\Phi,A)$ is 
realized that $\vp^A$ is realized, where the latter is obtained from 
$\vp$ by replacing each occurrence of a knowledge operator $K_i$ by $K_i^A$.)

In particular, in the single agent case, if we apply the synthesis
procedure of the previous section to the specification $\vp \And {\rm
Say}(\Phi)$, 
and then project away the ``saying'' component of the action function, 
we obtain a protocol that represents knowledge defined
according to the perfect-recall semantics, but using only a finite 
number of states.

\subsection{Implementing Knowledge Based Programs} \label{sec:kbp}

In the literature on knowledge-based specification of distributed
systems, many of the examples considered have the form of a
description of how an agent determines it next action from its state
of knowledge. To formalize this idea, Fagin et al. \cite{FHMVdc97,FHMV}
propose a syntax and semantics for what they call
\emph{knowledge-based programs}.  (The proposal builds on 
an earlier purely semantic framework of Halpern and Fagin
\cite{HF87}.) 
A knowledge-based program for an agent $i$ is an expression of the form 
\begin{equation} 
\begin{array}{l}
   \mbox{\bf case of}\\ ~~~~ {\bf if} ~ \vp^i_1 ~ {\bf do } ~ a^i_{1} \\
   ~~~~~~~ \vdots \\ ~~~~ {\bf if} ~ \vp^i_{m_i} ~ {\bf do } ~ a^i_{m_i} \\
   \mbox{\bf end case} \end{array}\label{eq:pg}
\end{equation} 
where the $\vp^i_j$ are formulae in the logic of knowledge that
express some property of agent $i$'s knowledge, and the $a^i_j$ are
actions of the agent. A \emph{joint} knowledge-based program consists
of such a program for each agent. To ensure that an agent is able
to determine whether the formulae in its program hold, these are
required to be \emph{local} to the agent. One way to ensure this
restriction is to require that the $\vp_j^i$ are boolean combinations
of formulae of the form $K_i \vp$.

Informally, such a program
represents an infinite loop.  At each point of time, the agent
determines which of the formulae $\vp_j^i$ hold, and
nondeterministically selects one of the correspond actions $a^i_j$ for
execution. Of course, in order to give semantics to the knowledge
formulae we require some interpreted system. This system is required
to be one that is obtained by running the program itself.

This description of the semantics of knowledge-based programs appears
circular, but it is not viciously so.  Fagin et al.~ show how to
eliminate the apparent circularity, by describing what it means for a
protocol (in the sense of the present paper), in which the agent's
actions are not dependent on it knowledge, but only upon properties of
its concrete state, to be an \emph{implementation} of a
knowledge-based program. Their definition uses a notion of
\emph{context}. Contexts have a structure similar to the environments
we have used in the present paper, and the semantics of
knowledge-based programs can also be stated in terms of environments,
as is done by van der Meyden \cite{meyden:tark96}.  We consider the
latter approach in the following.

We defined above the interpreted system $\cI(\bP,E)$ generated by a
(joint) protocol $\bP$ in an environment $E$. The idea underlying the
semantics of knowledge-based programs is that by interpreting the
tests for knowledge in a (joint) knowledge-based program $\pg$ with
respect to this system, we obtain for each agent, at each point of the
system, the set of actions enabled by the program. 
Formally, for each point $(r,m)$ of a system $\cI$, and agent $i$, 
we define $\pg^\cI_i(r,m)$ to be the set of actions $a^i_j$  such 
that $\cI,(r,m) \models \vp^i_j$. 
Similarly, the protocol $\bP$ prescribes a set of 
actions at each point of the system, computed from the agents' 
protocol states at the point. 
For agent $i$, this set is $\bP_i(r_i(m))$. 
We may now say that $\bP$ 
\emph{implements} $\pg$ in $E$ if these two sets of actions are
identical at every point of $\cI(\bP,E)$, 
i.e., $\pg^{\cI(\bP,E)}_i(r,m) = \bP_i(r_i(m))$ for all points $(r,m)$ of $\cI(\bP,E)$
and agents $i$.

A consequence of this definition of the semantics of knowledge-based
programs is that these may have one, many or no implementations. In
this regard, it has been noted that they are more like
\emph{specifications} than programs that may be written in a typical
imperative programming language. It is possible to make this intuition
precise using the definition of realizability of knowledge-based
specifications we have proposed in the present paper.  

\newcommand{\did}{{\rm did}}

Let $E=\la S_e, I_e, P_e, \tau , O_1,\ldots, O_n,\pi_e \ra$ be an
environment. We assume that that the set of basic propositions and the
set of states are sufficiently rich to express the latest joint action
taken in $E$. More precisely, we assume that for each agent $i$ and
for each action $a\in ACT_i$ there exists a proposition ``$\did_i(a)$'', and
that for all joint actions $\ba = \la a_e,a_1, \ldots, a_n\ra$ and
states $s\in S_e$, we have that if $\tau(\ba)(s) = t$ then
$\pi_e(t,\did_i(a)) = 1$ iff $a = a_i$. It is not difficult to see
that an environment can always be modified to one satisfying this
condition, by adding extra components to the states of the environment
in a fashion similar to the construction of the previous section.
We may then state the connection between knowledge-based programs 
and realizability  as follows: 

\pro
A joint protocol $\bP$ is an implementation (with respect to the
perfect-recall semantics of knowledge) of the joint knowledge-based
program, given by~(\ref{eq:pg}), in the environment $E$ iff $\bP$
realizes the specification
\[\forall \Box  \bigwedge_{i=1}^n  
\bigwedge_{j=1}^{m_i} (\vp^i_j \equiv \exists \Circ\  \did_i(a^i_j)) \] 
in $E$. 
\epro

\prf
The result follows straightforwardly from the facts that 
$\pg_i^{\cI(\bP,E)}(r,m) = \{a^i_j~|~\cI(\bP,E),(r,m) \models \vp^i_j\}$ and 
$\bP_i(r_i(m)) = \{a^i_j~|~\cI(\bP,E),(r,m) \models \exists \Circ\  \did_i(a^i_j))\}$
for all points $(r,m)$ of $\cI(\bP,E)$. 
\eprf

This relates a knowledge-based program for $n$ agents to a
CTL$^*$-K$_n$ formula.  It follows from this result that 
the techniques we have developed in this paper may be applied to 
construct an implementation of a knowledge-based program 
for a single agent, if one exists.

\section{Conclusion} \label{sec:concl}

We have been able to treat the case of single agent knowledge-based
specifications in this paper. Is it possible to generalize our results
to the multi-agent case? In general, it is not. Using ideas from
Peterson and Reif's study of the complexity of multi-agent games of
incomplete information \cite{PR79}, Pnueli and Rosner
\cite{PR90}  show that realizability of 
linear temporal logic specifications in
the context of two agents with incomplete information is undecidable.
This result immediately applies to our more expressive specification
language.

However, there are limited classes of situations in which
realizability of temporal specifications for more than one agent with
incomplete information is decidable, and for which one still obtains
finite-state implementations. Pnueli and Rosner
\cite{PR90} show that this is the case for \emph{hierarchical} agents. 
In our epistemic setting, a related (but not quite equivalent) notion is 
assumption that the
observation functions $O_i$ have the property that for all states $s$
and $t$ of the environment, if $O_i(s) = O_i(t)$ then $O_{i+1}(s) =
O_{i+1}(t)$.  Intuitively, this means that agent 1 makes more detailed
observations than agent 2, which in turn makes more detailed
observations than agent 3, etc. 
On the basis of Pnueli and Rosner's results, we conjectured in an earlier version of this paper
that realizability of
knowledge-based 
specifications in hierarchical
environments may also be decidable. 
This has subsequently been decided in the negative for the general case, but under some
extra conditions on the formula (that knowledge operators occur only positively), 
the conjecture turns out to be true \cite{MW05}. 

Other restrictions on the environment suggest themselves as candidates
for generalization of our results. For example, whereas atemporal
knowledge-based programs (in which conditions do not involve temporal
operators) do not have finite-state implementations in general
\cite{meyden:tark96},  in \emph{broadcast} environments this is guaranteed
\cite{meyden:fsttcs96}. Again, this suggests that realizability 
of knowledge specifications in broadcast environments is worth
investigation, particularly as this is a very natural and applicable
model. 
This conjecture has been shown to hold \cite{MW05}.

Finally, one could also consider definitions of knowledge other
than the perfect-recall interpretation that we have treated in this
paper. 
We believe that our techniques can be easily adapted to 
show the decidability of a synthesis for specifications concerning a single 
agent with asynchronous perfect recall. 
However, 
one open question is whether it is decidable to
determine the existence of a finite state automaton $A$ realizing a
specification stated using the knowledge operator $K^A_1$.  The
result of Section~\ref{sec:kii} provides only a sufficient condition
for this.

\bibliographystyle{plain}

{\small \bibliography{z,ok,synth}} 
\end{document}